\documentclass[lettersize,journal]{IEEEtran}
\usepackage{amsmath,amsfonts}
\usepackage{algorithmic}
\usepackage{array}
\usepackage[caption=false,font=normalsize,labelfont=sf,textfont=sf]{subfig}
\usepackage{textcomp}
\usepackage{stfloats}
\usepackage{url}
\usepackage{verbatim}
\usepackage{graphicx}
\usepackage{booktabs}
\usepackage{graphicx}
\usepackage{svg}
\usepackage{amssymb}
\usepackage{placeins}
\usepackage{bm}
\usepackage{booktabs}
\usepackage{multirow}
\usepackage{tabularray}

\def\BibTeX{{\rm B\kern-.05em{\sc i\kern-.025em b}\kern-.08em
    T\kern-.1667em\lower.7ex\hbox{E}\kern-.125emX}}
\usepackage{balance}
\begin{document}
\title{FD-LSCIC: Frequency Decomposition-based Learned Screen Content Image Compression}
\author{Shiqi Jiang, Hui Yuan, Senior Member, IEEE, Shuai Li, Senior Member, IEEE, Huanqiang Zeng, and Sam Kwong, Fellow, IEEE
\thanks{This work was supported in part by the National Natural Science Foundation of China under Grants 62222110 and 62172259, the Taishan Scholar Project of Shandong Province (tsqn202103001), the Natural Science Foundation of Shandong Province under Grant ZR2022ZD38. (Corresponding author: Hui Yuan)

Shiqi Jiang and Shuai Li are with the School of Control Science and Engineering, Shandong University, Ji'nan, 250100, China. (Email: shiqijiang@mail.sdu.edu.cn, shuaili@sdu.edu.cn)

Hui Yuan is with the School of Control Science and Engineering, Shandong University, Ji'nan, 250100, China, and also with the Shandong Inspur Artificial Intelligence Research Institute Co., Ltd., Ji'nan, China.(Email: huiyuan@sdu.edu.cn)

Huanqiang Zeng is with the School of Information Science and Engineering,
Huaqiao University, Xiamen 361021, China (e-mail: zeng0043@hqu.edu.cn).

Sam Kwong is with the School of Data Science, Lingnan University, Hong Kong (e-mail: samkwong@ln.edu.hk).
}}

\markboth{Journal of \LaTeX\ Class Files,~Vol.~18, No.~9, September~2020}%
{How to Use the IEEEtran \LaTeX \ Templates}

\maketitle

\begin{abstract}
The learned image compression (LIC) methods have already surpassed traditional techniques in compressing natural scene (NS) images. However, directly applying these methods to screen content (SC) images, which possess distinct characteristics such as sharp edges, repetitive patterns, embedded text and graphics, yields suboptimal results. This paper addresses three key challenges in SC image compression: learning compact latent features, adapting quantization step sizes, and the lack of large SC datasets. To overcome these challenges, we propose a novel compression method that employs a multi-frequency two-stage octave residual block (MToRB) for feature extraction, a cascaded triple-scale feature fusion residual block (CTSFRB) for multi-scale feature integration and a multi-frequency context interaction module (MFCIM) to reduce inter-frequency correlations. Additionally, we introduce an adaptive quantization module that learns scaled uniform noise for each frequency component, enabling flexible control over quantization granularity. Furthermore, we construct a large SC image compression dataset (SDU-SCICD10K), which includes over 10,000 images spanning basic SC images, computer-rendered images, and mixed NS and SC images from both PC and mobile platforms. Experimental results demonstrate that our approach significantly improves SC image compression performance, outperforming traditional standards and state-of-the-art learning-based methods in terms of peak signal-to-noise ratio (PSNR) and multi-scale structural similarity (MS-SSIM). The source code will be opened later in https://github.com/SunshineSki/Screen-content-image-dataset/tree/main/SDU-SCICD10K.
\end{abstract}


\section{Introduction}
The surge in digital devices and the increasing reliance on screen-based activities have led to a significant rise in the generation and consumption of screen content (SC). From remote work environments and online education to gaming and digital media creation, SC now constitutes a substantial portion of the visual data transmitted and stored globally. This growth has introduced new challenges in effectively processing and compressing SC. Traditional video coding standards, evolving from JPEG \cite{wallace1991jpeg} to H.265/HEVC \cite{pan2016fastmotion} and H.266/VVC \cite{bross2021overview}, have achieved significant improvements in compression performance, primarily optimized for natural scene (NS) images through the incorporation of handcrafted priors. In recent years, learned image compression (LIC) methods \cite{Zou_2022_CVPR,liu2023learned} have surpassed even the latest coding standard, i.e., H.266/VVC, in terms of coding performance. However, these advancements are derived from complex network models jointly trained on large NS image datasets, and their direct application to SC image compression yields disappointing results. This is because SC images, generated by computers, fundamentally differ from NS images captured by cameras, possessing unique attributes such as sharp edges, repetitive patterns \cite{zhang2023subjective}. As the data volume of SC on internet continues to increase rapidly, there is an urgent need for advanced compression techniques specifically designed to address the distinct characteristics of this type of data.
\begin{figure}[t]
	\centering
	\begin{minipage}{0.5\textwidth}
	\centering
	\includegraphics[width=\textwidth]{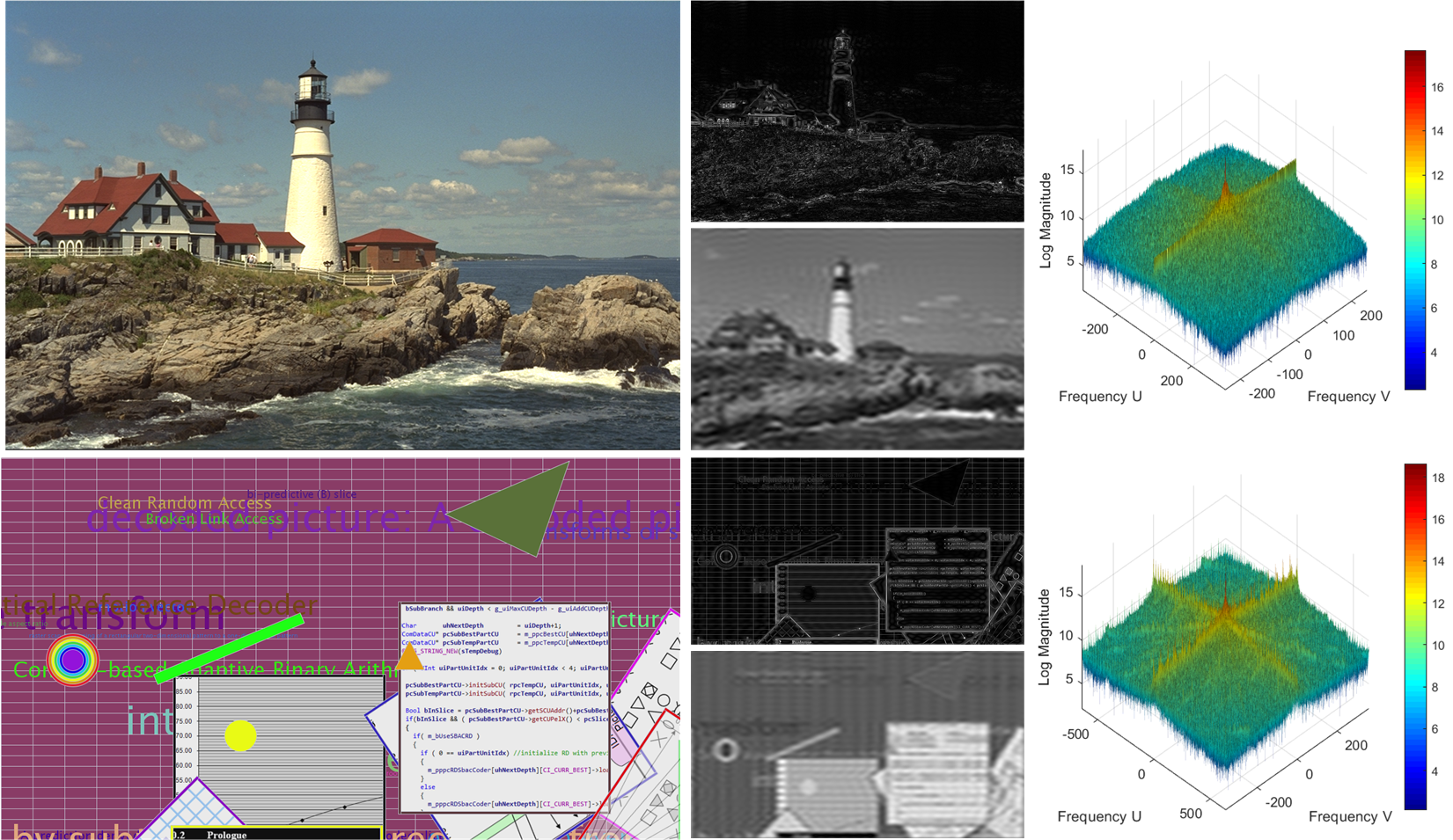}
	\caption{Frequency characteristics of NS and SC images. The left column represents the original image, the middle column represents their corresponding high-frequency and low-frequency decompositions, and the right column represents the frequency spectrum distribution.}
	\label{fig:Frequency characteristics}
	\end{minipage}
    \vspace{-0.4cm}
\end{figure}

The excellent performance of LIC methods for NS image compression highlights its vast potential for SC image compression. LIC methods formulate image compression as a coding problem of nonlinear transformations, with the process comprising modules such as transformation, quantization, and entropy coding. The first critical issue is learning more compact latent features. For NS images, techniques like deeper networks \cite{fu2023learned}, attention model \cite{cheng2020learned}, invertible transformation \cite{xie2021enhanced}, and transformers \cite{liu2023learned,wangFDnet} are typically introduced to enhance feature learning, but these approaches often overlook the inherent characteristics of the SC images. As shown in Fig. \ref{fig:Frequency characteristics}, the transition between high and low frequencies is smoother in NS images, whereas SC images typically exhibits more extreme high and low-frequency components. Therefore, leveraging frequency distribution characteristics for SC image compression is a feasible approach that differentiates from NS image compression. The second critical issue is the quantization of latent features. Most existing methods use a fixed quantization step size (typically set to 1) for quantization. Guo {\em et al.} \cite{guo2021soft} introduced a scaled uniform noise whose scale can be learnt to adapt the quantization step size by deriving a new variational upper bound on the actual bitrate. However, this method only considers the noise interval parameter of single-frequency features and does not allow flexible control over the quantization granularity of different frequency components. The third critical issue is the lack of large publicly available datasets for SC image compression. Existing SC datasets are not large enough for learning-based compression \cite{jiang2024OMR} and are primarily focused on quality assessment \cite{wang2021quality}. Consequently, for data-driven SC image compression, the limited training data constrains the rate-distortion (RD) performance.

To address the three key issues mentioned above, we propose a multi-frequency decomposition and adaptive quantization method specifically designed for SC image compression. We leverage the distribution characteristics of SC images by dividing the image features into high, mid, and low-frequency components. Building on the multi-frequency decomposition, different noise interval parameters are then learned for each feature component to adjust the corresponding quantization step size, enabling more flexible quantization granularity to enhance coding performance. In addition, we construct a large screen content image compression dataset (SDU-SCICD10K), which is categorized into three types: web and office images, computer-rendered images, and mixed NS and SC images, with a total of over 10,000 images. In detail, our contributions are listed as follows.

\begin{itemize}
	\item We propose a compression method for SC images that performs feature extraction using a multi-frequency two-stage octave residual block (MToRB) to capture high, mid, and low-frequency components, a cascaded triple-scale feature fusion residual blocks (CTSFRB) to integrate features across different scales and a multi-frequency context interaction module to reduce inter-frequency correlations.
 
    \item We propose an adaptive quantization module tailored for different frequency components, which adjusts the quantization step size by learning the scaled uniform noise for each frequency component, enabling more flexible control over quantization granularity.

	\item We propose a large screen content image compression dataset (SDU-SCICD10K), which is broadly categorized into three types, i.e., web and office images, computer-rendered images, and mixed NS and SC images, each containing various image styles from both PC and mobile platforms.

	\item Experimental results demonstrate the effectiveness of the proposed method in improving the RD performance of SC image compression. Our approach outperforms traditional image compression standards, including the state-of-the-art H.266/VVC with screen content coding (SCC) techniques (H.266/VVC-SCC) and state-of-the-art learning-based methods in terms of both PSNR and MS-SSIM at the same coding bitrate.
\end{itemize}

The remainder of this paper is organized as follows. In section II, we briefly review related work on SC image compression and the octave convolution that is used in the proposed method. In section III, we formulate the compression problem mathematically to guide the design of the proposed method. Then, in section IV, the proposed method is described in detail. Experimental results and conclusions are given in Sections V and VI, respectively.
\begin{figure}[t]
\begin{minipage}[t]{0.3\linewidth}
  \centering
  \includegraphics[width=0.9\linewidth]{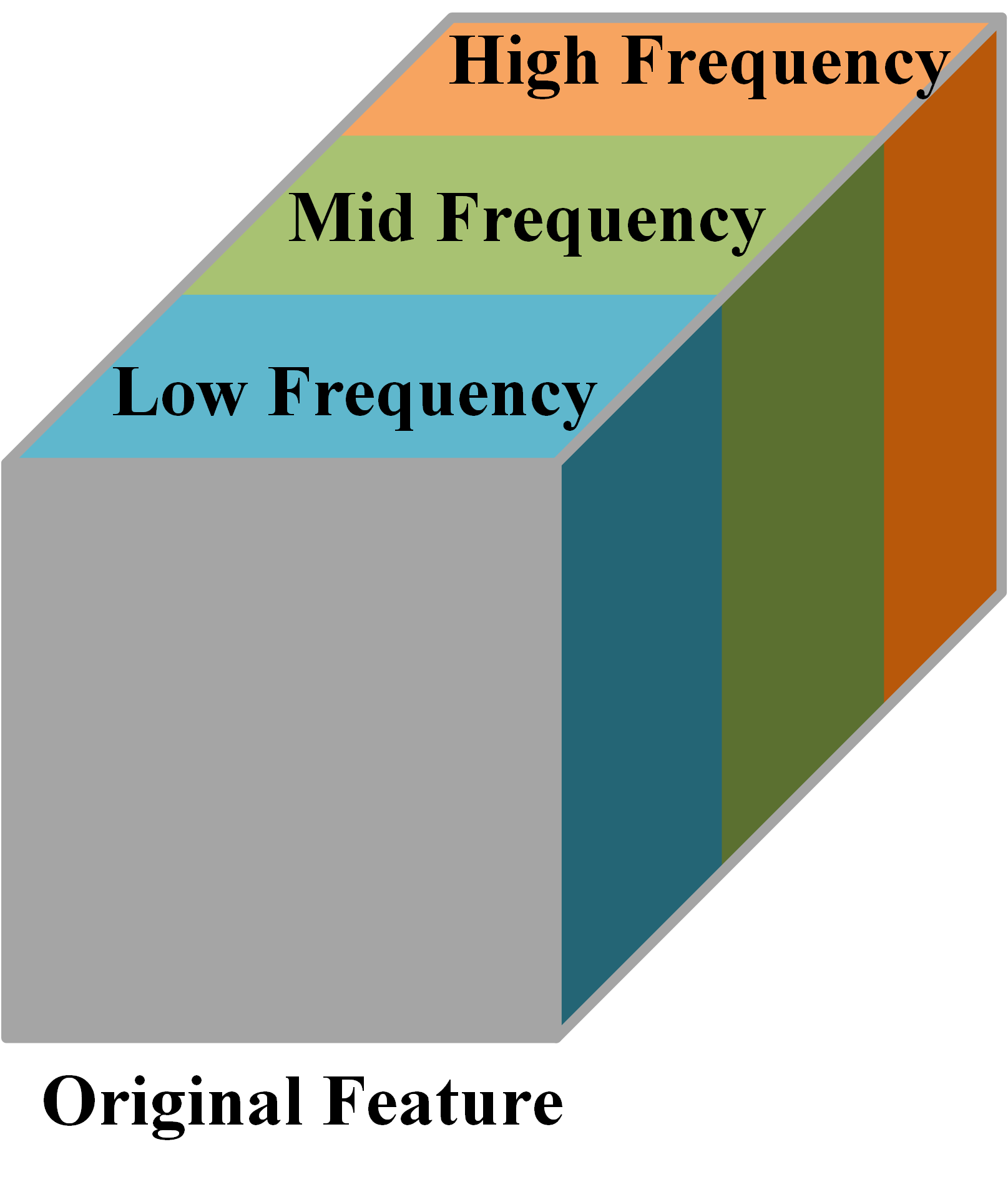}
  \centerline{(a)}\medskip
\end{minipage}
\begin{minipage}[t]{0.3\linewidth}
  \centering
  \includegraphics[width=0.9\linewidth]{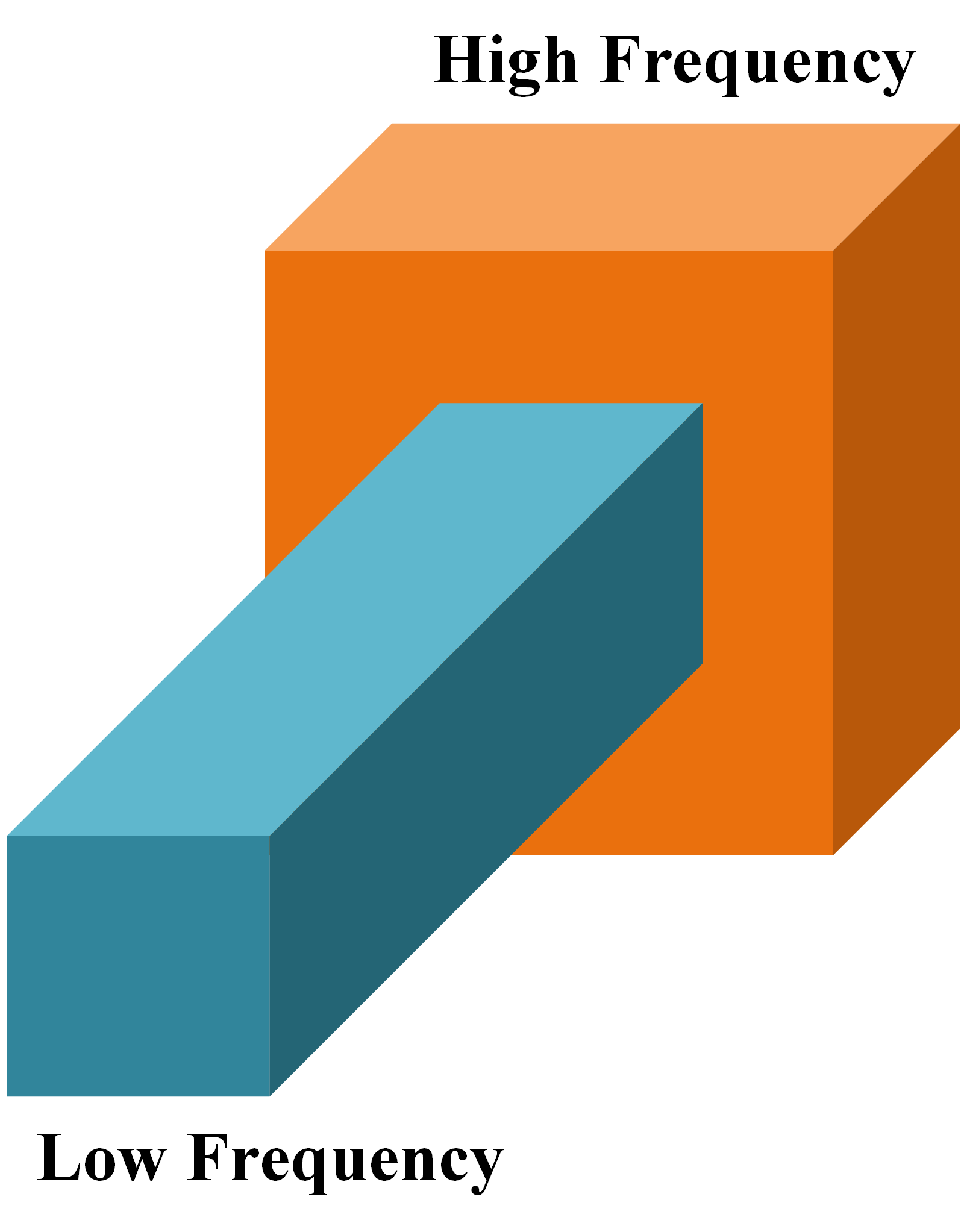}
  \centerline{(b)}\medskip
\end{minipage}
\begin{minipage}[t]{0.3\linewidth}
  \centering
  \includegraphics[width=0.9\linewidth]{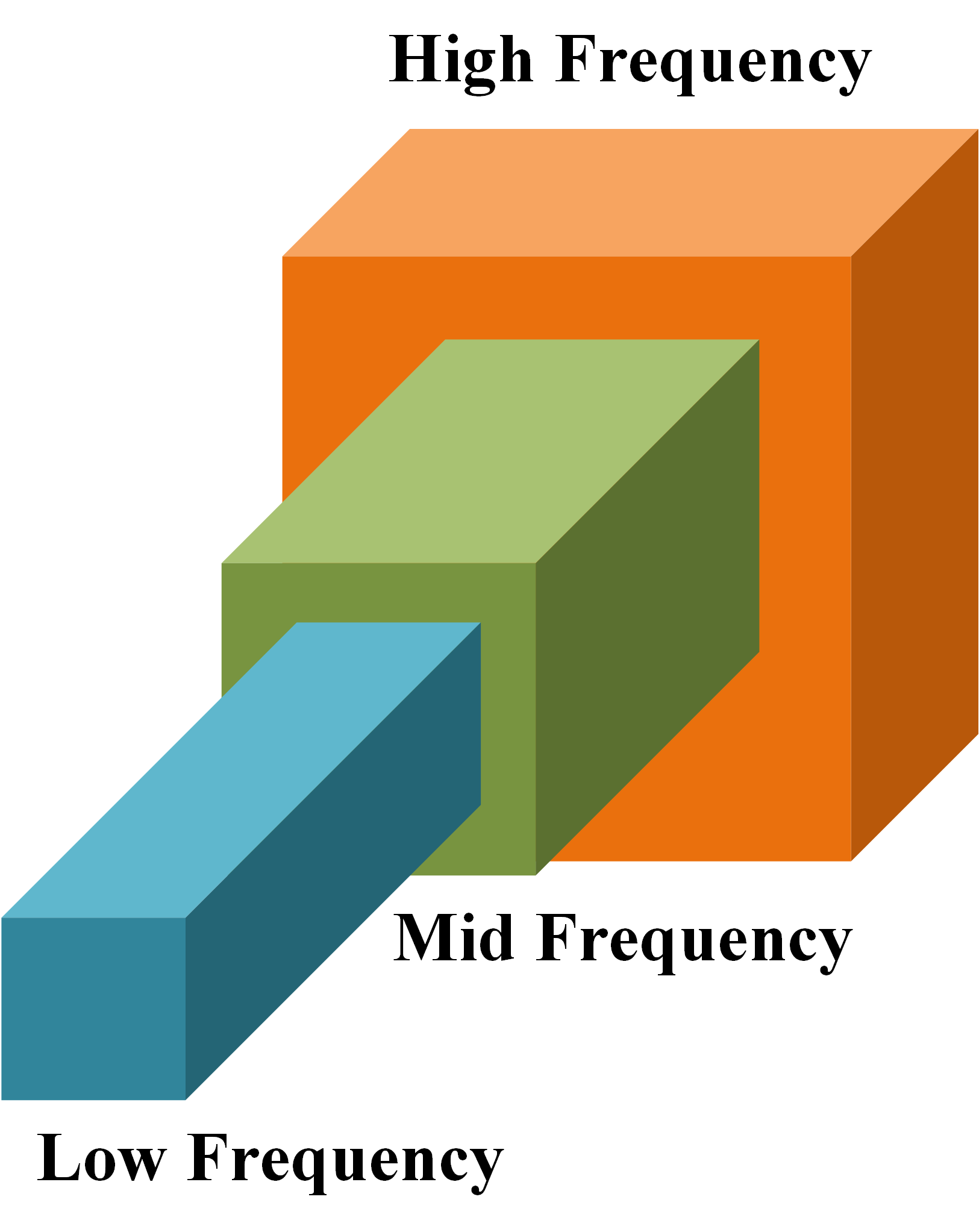}
  \centerline{(c)}\medskip
\end{minipage}
\begin{minipage}[t]{\linewidth}
  \centering
  \includegraphics[width=0.7\linewidth]{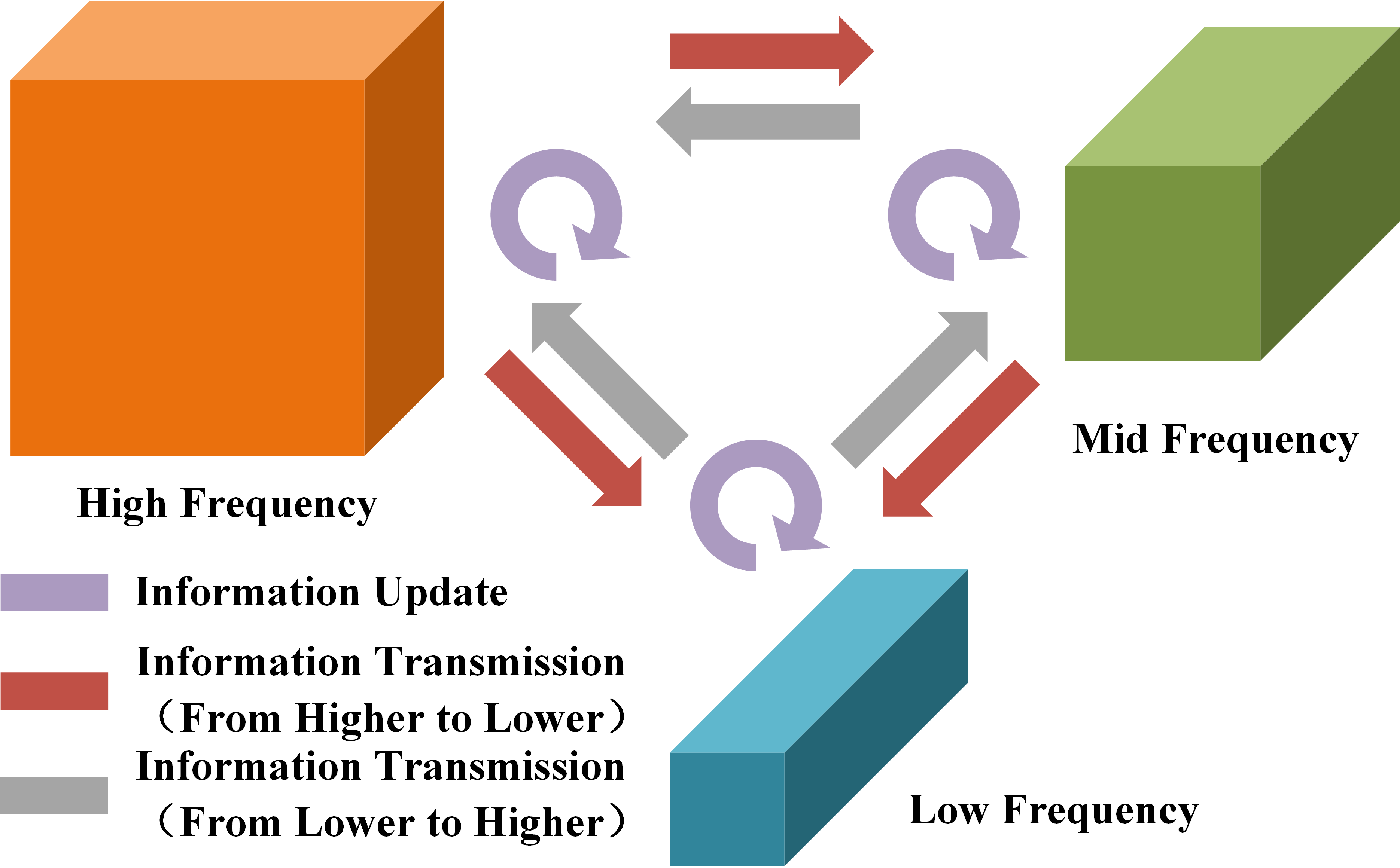}
  \centerline{(d)}\medskip
\end{minipage}
\caption{Illustration of OctConv-based frequency decomposition. (a) The original image features can be decomposed into a combination of multiple frequency components. (b) OctConv simplifies this by dividing the features into two parts: high-frequency in high-resolution tensors and low-frequency in low-resolution tensors. (c) The proposed multi-frequency octave convolution adds a middle-frequency component, where smoothly varying features are stored in lower-resolution tensors, sharply varying features in higher-resolution tensors, and the remaining features in intermediate-resolution tensors. (d) Different frequency components undergo intra-frequency information updates (purple arrows), information transfer from higher to lower frequencies (red arrows), and information transfer from lower to higher frequencies (gray arrows).}
\label{fig:motivation}
\end{figure}

\section{Related works}
\subsection{Screen content coding}
Traditional video coding methods, such as H.265/HEVC \cite{Lei2017fastintra}, H.266/VVC \cite{bross2021overview}, AVS3 \cite{yin2024afastcu}, and AV1 \cite{Hao2024FastTransform}, have increasingly focused on SCC by introducing specialized SCC tools to enhance encoding efficiency for SC \cite{zhu2017inter-palette,yang2020anovel,zhao2024gcostc}. In H.265/HEVC-SCC, techniques such as intra block copy (IBC) \cite{xu2016intra}, palette mode (PLT) \cite{pu2016palette}, adaptive color transform (ACT) \cite{zhang2016adaptive}, and transform skip mode (TSM) \cite{sole2013ahg8} were introduced. H.266/VVC-SCC \cite{wang2024fastmode} further improves them by incorporating transform skip residual coding (TSRC) and block-based differential pulse-code modulation (BDPCM) \cite{abdoli2019intra}. Similarly, AVS3-SCC introduces frequency-based intra-mode coding (FIMC) \cite{li2019freq}, implicit selection of transform skip (ISTS) \cite{zhang2021implicit}, adaptive control of deblocking type (ACDT) \cite{wang2020deblocking}, and intra string copy (ISC) \cite{zhou2020string} techniques.
\begin{figure}[t]
\centering
\begin{minipage}[b]{0.48\linewidth}
  \centering
  \includegraphics[width=1\linewidth]{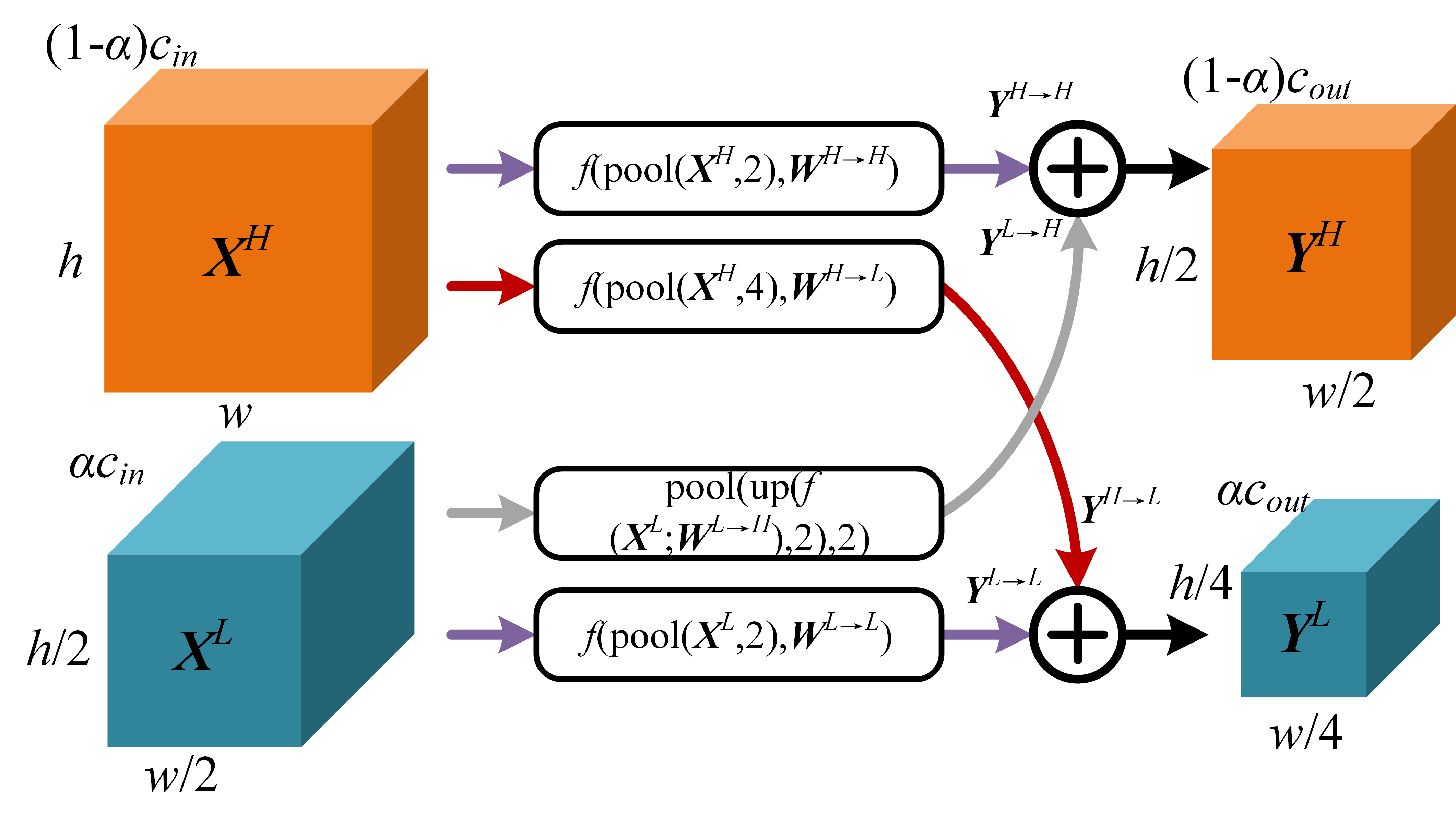}
  \centerline{(a) OctConv}\medskip
\end{minipage}
\begin{minipage}[b]{0.48\linewidth}
  \centering
  \includegraphics[width=1\linewidth]{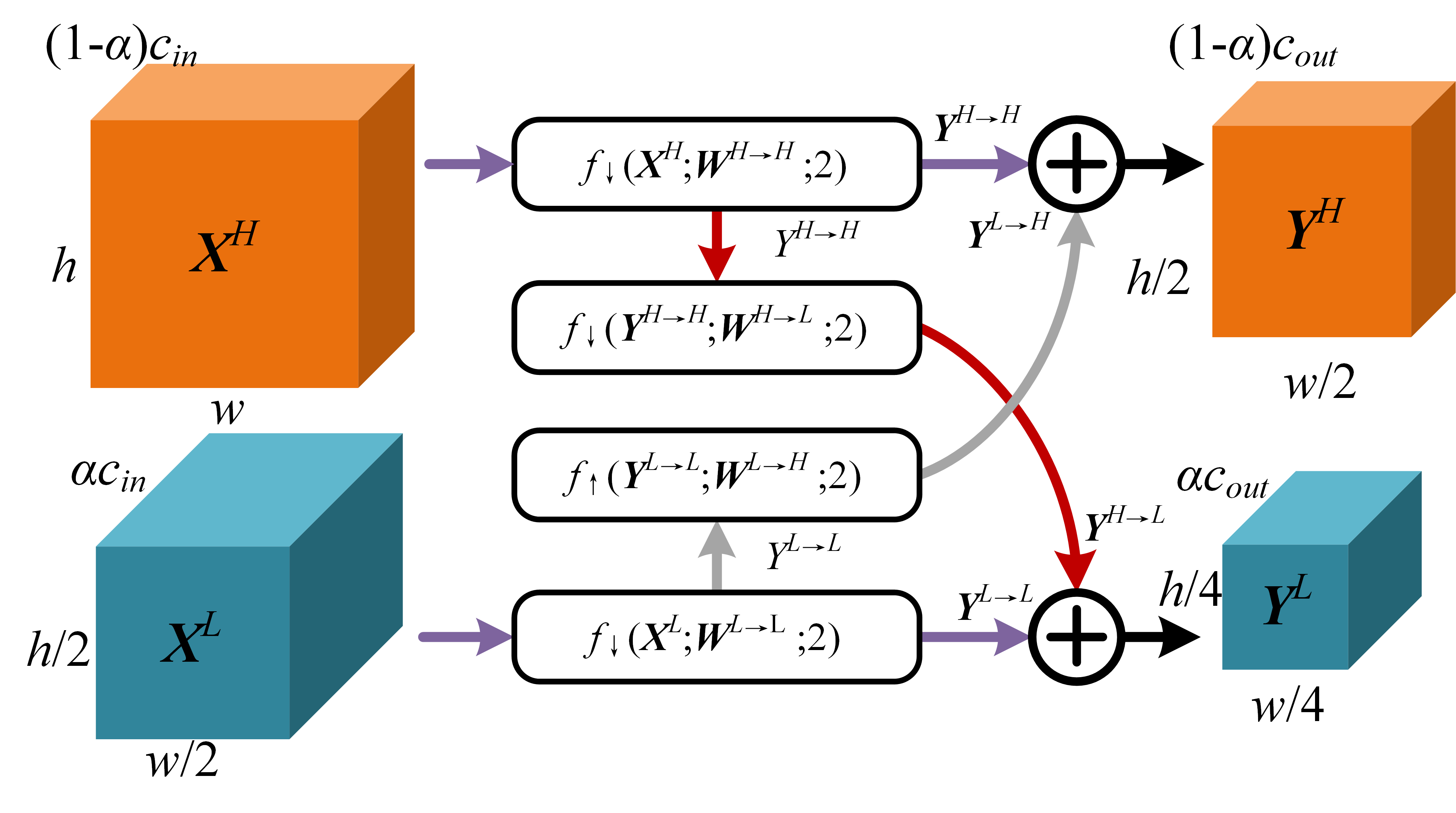}
  \centerline{(b) GoConv}\medskip
\end{minipage}
\begin{minipage}[b]{\linewidth}
  \centering
  \includegraphics[width=0.9\linewidth]{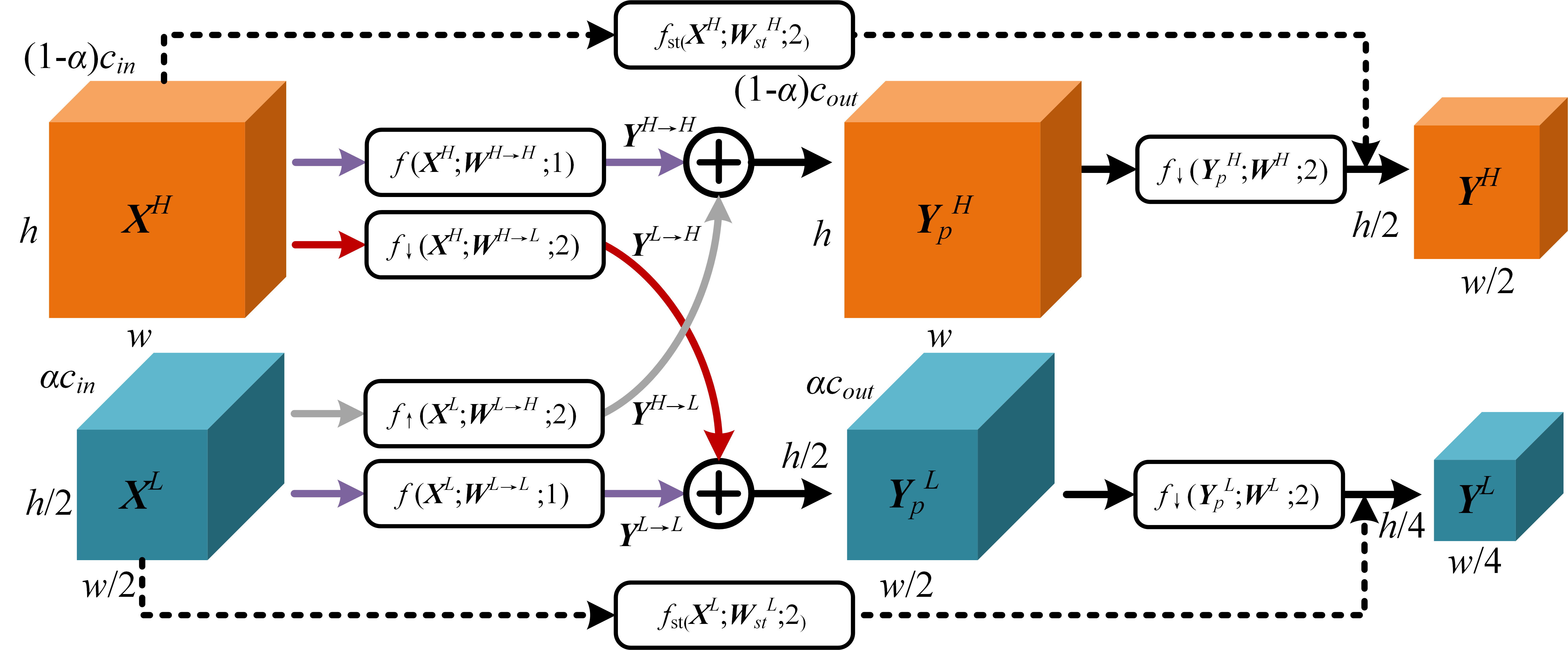}
  \centerline{(c) ToRB}\medskip
\end{minipage}
\begin{minipage}[b]{\linewidth}
  \centering
  \includegraphics[width=1\linewidth]{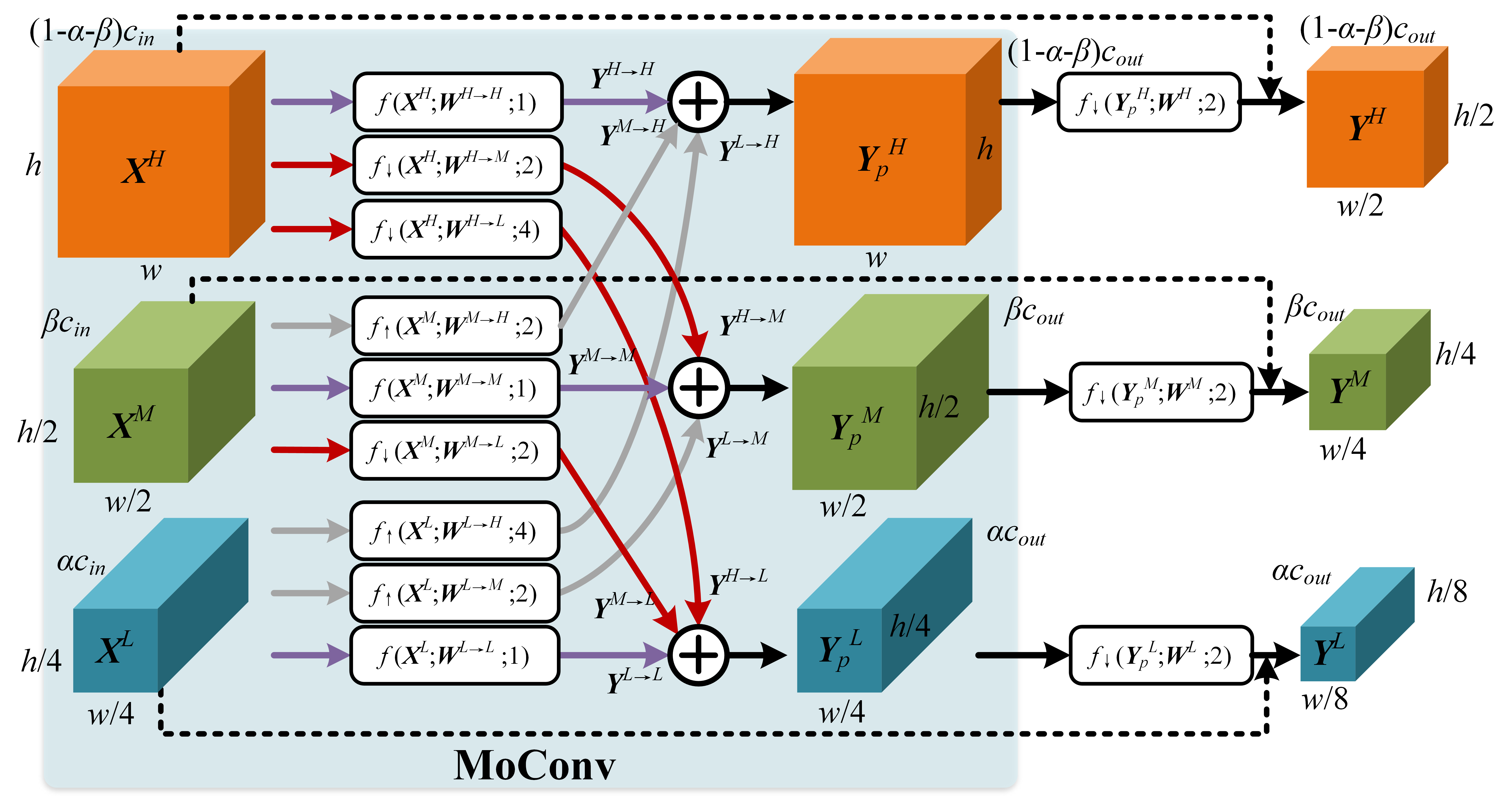}
  \centerline{(d) MToRB}\medskip
\end{minipage}
\caption{Down-sampling octave convolution and its variants, (a) OctConv \cite{chen2019drop}, (b) GoConv \cite{akbari2021learned}, (c) ToRB \cite{chen2022two}, and (d) MToRB (proposed). \({\alpha}\) denotes the channel ratio allocated to low-frequency features, \({\beta}\) denotes the channel ratio allocated to mid-frequency features, and the channel ratio for high-frequency features is \(1 - {\alpha} - {\beta}\).}
\label{fig:octave methods}
\end{figure}
With the rapid advancements in deep learning and the availability of large dataset, LIC methods have surpassed traditional techniques in compressing NS images \cite{cheng2020learned, balle2018variational, minnen2018joint, pan2024JND-LIC, he2024learnedimage}. These advancements have also begun to demonstrate significant potential in the domain of SC image compression \cite{wang2022transform, heris2023multi, shen2023dec, tang2023feature, WangDSCIC}. Wang \textit{et al}. \cite{wang2022transform} enhanced SC encoding performance by integrating transform skip into the framework presented in \cite{balle2018variational}. Heris \textit{et al}. \cite{heris2023multi} trained an encoder for both reconstruction and segmentation tasks, segmenting SC into synthetic and natural regions to meet multi-task encoding requirements. Shen \textit{et al}. \cite{shen2023dec} proposed an efficient decoder-side adapter for adaptive compression of both NS and SC images. Tang \textit{et al}. \cite{tang2023feature} addressed low-bitrate SCC method by employing a super-resolution network to enhance feature during the fusion process, thereby reducing compression distortion. Wang \textit{et al}. \cite{WangDSCIC} proposed a color context generator (CCG) and a region-based block aggregation (RBA) module to investigate color representations
and block correlations for SCC. Although these methods attempt to enhance SCC performance using learning-based techniques, they do not fully consider the inherent spectrum distribution characteristics of SC. While our previous research \cite{jiang2024OMR} consideres differences in spectrum distribution, it only accounts for two components: high-frequency and low-frequency. Furthermore, applying the same quantization operation to different components is not in line with the characteristic of human visual system \cite{pointer1989contrast}. Moreover, another limitation of SCC lies in the dataset. Although we constructed the SDU-SCICK2K dataset in \cite{jiang2024OMR}, it is still small that leaves significant room for improving SCC performance.

\subsection{Octave convolution}
Inspired by the frequency decomposition of images, Chen \textit{et al}. \cite{chen2019drop} proposed the octave convolution network (OctConv) to decompose features. It has demonstrated superior performance in various downstream tasks, such as image classification and action recognition. Besides, its ability to decompose features into different frequencies and reduce spatial redundancy is also well-suited for image compression \cite{jiang2024OMR}, indicating its broad potential in the encoding domain. Fig. \ref{fig:octave methods} (a) illustrates the structure of the original OctConv. The input features \(\bm{X}\) are decomposed into high-frequency part \(\bm{X}^H\) and low-frequency part \(\bm{X}^L\) along the channel dimension. They are then processed separately through intra-frequency convolution and inter-frequency convolution operations, resulting in intra-frequency features \(\{ \bm{Y}^{H \rightarrow H}, \bm{Y}^{L \rightarrow L} \}\) and inter-frequency features \(\{ \bm{Y}^{H \rightarrow L}, \bm{Y}^{L \rightarrow H} \}\). The final output consists of high-frequency features \(\bm{Y}^H\) and low-frequency features \(\bm{Y}^L\):
\begin{align}
\bm{Y}^H &= \bm{Y}^{H \rightarrow H} + \bm{Y}^{L \rightarrow H} = f\left(\operatorname{pool}\left(\bm{X}^H, 2\right); \bm{W}^{H \rightarrow H}\right) \nonumber \\
& \quad + \operatorname{pool}\left(up\left(f\left(\bm{X}^L \cdot \bm{W}^{L \rightarrow H}\right); 2\right); 2\right), \\
\bm{Y}^L &= \bm{Y}^{L \rightarrow L} + \bm{Y}^{H \rightarrow L} = f_{\text{pool}}\left(\bm{X}^L; \bm{W}^{L \rightarrow L}; 2\right) \nonumber \\
& \quad + f\left(\operatorname{pool}\left(\bm{X}^H, 4\right); \bm{W}^{H \rightarrow L}\right) ,
\end{align}
where \( f(\bm{X}; \bm{W}) \) denotes a convolution operation with parameters \(\bm{W} \), \( \text{pool}(\bm{X}; a) \) represents an average pooling operation with a stride of \( a \), and \( \text{up}(\bm{X}; b) \) represents an upsampling interpolation with a scale factor of \( b \).

Akbari \textit{et al}. \cite{akbari2021learned} observed that directly applying pooling operations with large strides in OctConv results in the loss of spatial information. To address this problem, they proposed a generalized octave convolution (GoConv) for image compression, based on OctConv, as shown in Fig. \ref{fig:octave methods} (b). The key differences include replacing the pooling operation with strided convolution and modifying the input of the inter-frequency convolution from \(\bm{Y}^H\) and \(\bm{Y}^L\) to the inter-frequency features \(\bm{Y}^{H \rightarrow H}\) and \(\bm{Y}^ {L \rightarrow L}\):
\begin{equation}
\bm{Y}^H = f_{\downarrow}\left(\bm{X}^H; \bm{W}^{H \rightarrow H}; 2\right) + f_{\uparrow}\left(\bm{Y}^{L \rightarrow L}; \bm{W}^{L \rightarrow H}; 2\right),
\end{equation}
\begin{equation}
\bm{Y}^L = f_{\downarrow}\left(\bm{X}^L; \bm{W}^{L \rightarrow L}; 2\right) + f_{\downarrow}\left(\bm{Y}^{H \rightarrow H}; \bm{W}^{H \rightarrow L}; 2\right),
\end{equation}
where $f_{\downarrow}(\bm{X};\bm{W};a)$ denotes downsampling convolution with parameters $\bm{W}$ and a stride of \( a \), $f_{\uparrow}(\bm{X};\bm{W};b)$ denotes upsampling convolution with parameters $\bm{W}$ and a stride of \( b \).

Chen \textit{et al}. \cite{chen2022two} found that, in GoConv, applying \(\bm{Y}^{H \rightarrow H}\) and \(\bm{Y}^{L \rightarrow L}\) to both intra-frequency and inter-frequency convolutions simultaneously would limit the neural network's ability to learn different frequency features effectively. Additionally, the downsampling of \(\bm{X}^L\) followed by upsampling increases distortion. To address these issues, they proposed a two-stage octave residual block (ToRB), as shown in Fig. \ref{fig:octave methods} (c), which separates the sampling operations:
\begin{flalign}
&\begin{aligned}
\bm{Y}^H & =f_{\downarrow}\left(\bm{Y}^H_p; \bm{W}^H ; 2\right)+f_{\mathrm{sc}}\left(\bm{X}^H; \bm{W}_{sc}^H ; 2\right), 
\end{aligned}\\
&\begin{aligned}
\textit{with } \bm{Y}^H_p & =\bm{Y}^{H \rightarrow H} + \bm{Y}^{L \rightarrow H} \\
& =f\left(\bm{X}^H; \bm{W}^{H \rightarrow H}\right)+f_{\uparrow}\left(\bm{X}^L;\bm{W}^{L \rightarrow H} ; 2\right),
\end{aligned}\\
&\begin{aligned}
\bm{Y}^L & =f_{\downarrow}\left(\bm{Y}_p^L; \bm{W}^L; 2\right)+f_{\mathrm{sc}}\left(\bm{X}^L;\bm{W}_{sc}^L; 2\right), 
\end{aligned}&\\
&\begin{aligned}
\textit{with }  \bm{Y}^L_p & =\bm{Y}^{L \rightarrow L} + \bm{Y}^{H \rightarrow L} \\
& =f\left(\bm{X}^L; \bm{W}^{L \rightarrow L}\right)+f_{\downarrow}\left(\bm{X}^H; \bm{W}^{H \rightarrow L}; 2\right),
\end{aligned}
\end{flalign}
where \(\bm{Y}^H_p\) and \(\bm{Y}^L_p\) represent the high- and low-frequency features of the first stage, respectively, and \(f_{sc}(\bm{X}; \bm{W}_{sc})\) denotes the skip connection with parameters \(\bm{W}_{sc} \).

Although OctConv-based methods have been continuously improved and successfully applied to image compression with commendable results, there still remains potential for enhancement in SCC. Therefore, considering the characteristics of SC images, we propose a multi-frequency feature  decomposition network to achieve more refined feature extraction and processing, as shown in Fig. \ref{fig:motivation}. The specific details will be elaborated in Section. IV.
\section{Preliminary and problem formulation}
\begin{figure}[t]
	\centering
	\begin{minipage}{0.5\textwidth}
	\centering
	\includegraphics[width=0.9\textwidth]{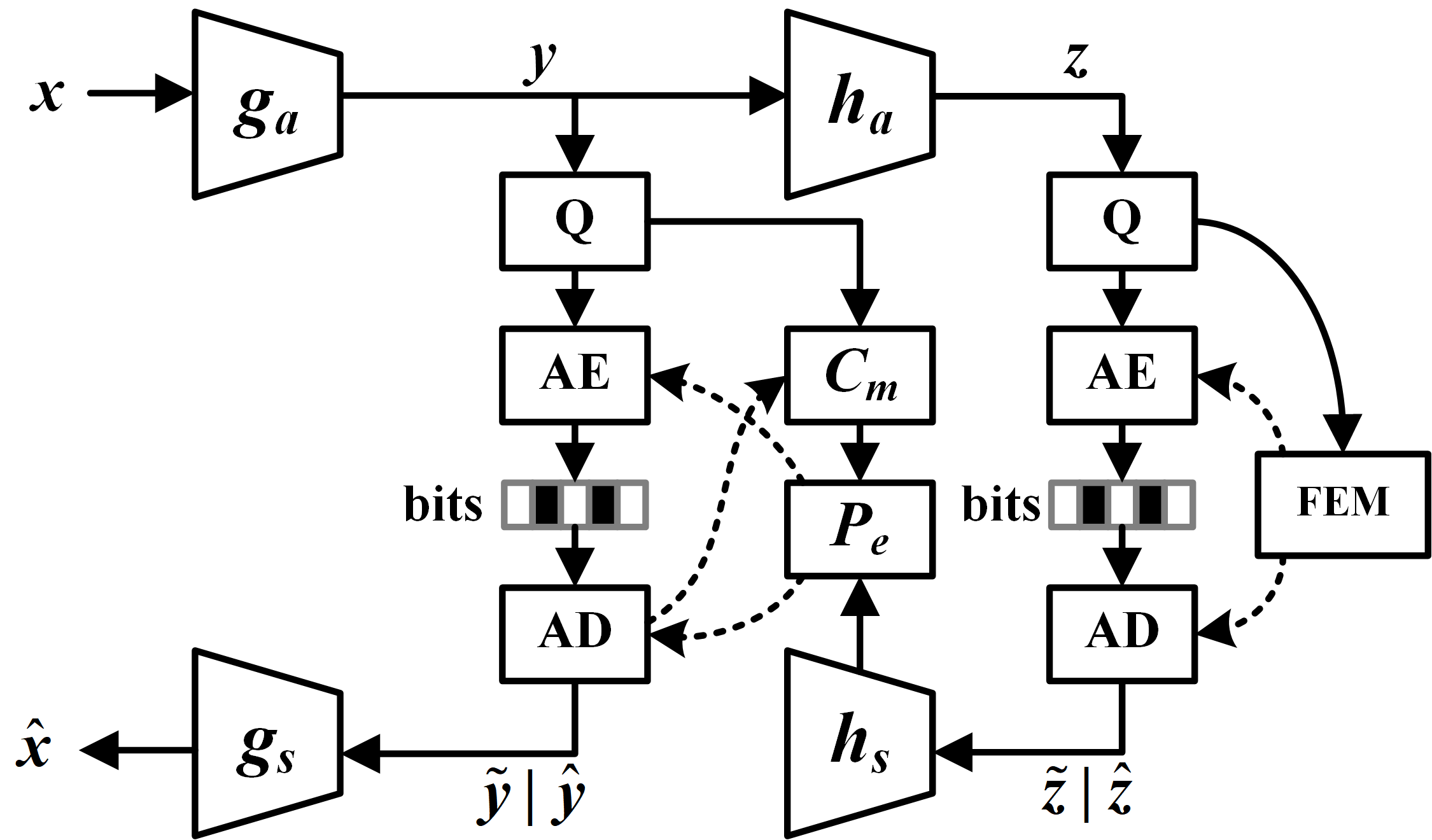}
	\caption{Framework of LIC methods. $g_a$ and $g_s$ denote main encoder and decoder, respectively. $h_a$ and $h_s$ denote hyper encoder and decoder, respectively. $C_m$ denotes context model, $P_e$ denotes entropy parameter model, FEM denotes the factorized entropy model \cite{balle2018variational}, $\mathrm{Q}$ denotes quantization module. The latent feature $\bm{y}$ (resp. hyper latent feature $\bm{z}$) is quantized as $\bm{\tilde{y}}$ (resp. $\bm{\tilde{z}}$) for training and $\bm{\hat{y}}$ (resp. $\bm{\hat{z}}$) for testing. AE and AD denote arithmetic encoding and arithmetic decoding, respectively.}
	\label{fig:LIC framework}
	\end{minipage}
\end{figure}
The LIC framework can be described as an optimization problem based on nonlinear transformations \cite{balle2017end}, as shown in Fig. \ref{fig:LIC framework} (left side). First, the analysis transformations (main encoder) $g_a$ takes the input image $\bm{x}$ and outputs the latent feature representation $\bm{\hat{y}}$:
\begin{equation}
\bm{y} = g_a(\bm{x}).
\end{equation}

Then, $\bm{y}$ can be quantized to achieve high compression efficiency. However, the quantization operation $Q$ is non-differentiable and thus hard to be trained by backward propagation. To address this problem, soft quantization techniques are employed. Specifically, in the training stage, $\bm{y}$ is soft quantized by adding uniform noise $\bm{n}$ to get the soft quantized latent feature $\bm{\tilde{y}}$, while in the testing stage, $\bm{y}$ is hard quantized by rounding operation to get the hard quantized latent feature $\bm{\hat{y}}$:
\begin{align}
\bm{\tilde{y}} &= \bm{y}+\bm{n}, \bm{n}\in U(\frac{1}{2},\frac{1}{2}),\\
\bm{\hat{y}} &= Q(\bm{y})=round(\bm{y}). 
\end{align}

For quantized features, entropy coding is used to generate a code stream under a given probabilistic model. The reconstructed image $\bm{\hat{x}}$ is obtained by the synthesis transformation (main decoder) $g_s$:
\begin{equation}
\bm{\hat{x}}=g_s(\bm{\hat{y}}).
\end{equation}

To dig the intra-correlation in $\bm{y}$ efficiently, later work \cite{balle2018variational,minnen2018joint} introduce hyper latent feature and context model based on the variational autoencoder (VAE) to construct a more flexible entropy model, as shown in the right side of Fig. \ref{fig:LIC framework}. An additional feature $\bm{z}$ is introduced to capture spatial correlations, obtained through the hyper analysis transformation (hyper encoder) $h_a$:
\begin{equation}
\bm{z} = h_a(\bm{y}).
\end{equation}

The hyper latent feature $\bm{z}$ should also be quantized by adding uniform noise $\bm{n}$ to obtain the quantized hyper latent feature $\bm{\tilde{z}}$. The density distribution $p_{\bm{y}}$ of the latent feature $\bm{y}$, which follows normalized distribution with parameters $\bm{\mu}$ and $\bm{\sigma}$, can be obtained using an entropy parameter model $P_e$ and a hyper synthesis transformation $h_s$ applied to $\bm{\tilde{z}}$:
\begin{equation}
(\bm{\mu},\bm{\sigma}) = P_{e}(h_s(\bm{\tilde{z}})).
\end{equation}

As the quantization operation can be achieved by adding noise into the latent feature during training, the conditional density distribution of quantized latent feature, i.e.,  $p_{\bm{\tilde{y}} \mid \bm{\tilde{z}}}$, can be obtained by convolving the density model $p_{\bm{y}}$ with the standard uniform distribution \cite{balle2018variational}:
\begin{equation}
p_{\bm{\tilde{y}} \mid \bm{\tilde{z}}}(\bm{\tilde{y}} \mid \bm{\tilde{z}})=\prod_i\left(\mathcal{N}\left(\mu_i, \sigma_i\right) * \mathcal{U}\left(-\frac{1}{2}, \frac{1}{2}\right)\right)\left(\tilde{y}_i\right),
\label{eq:density}
\end{equation}
in which hyper latent feature $\bm{z}$ is modeled by a fully factorized entropy model:
\begin{equation}
p_{\bm{\tilde{z}} \mid \bm{\Theta}}(\bm{\tilde{z}} \mid \bm{\Theta})=\prod_i\left(p_{z_i \mid \bm{\Theta}}(\bm{\Theta}) * \mathcal{U}\left(-\frac{1}{2}, \frac{1}{2}\right)\right)\left(\tilde{z}_i\right),
\label{eq:factorized}
\end{equation}
where $\bm{\Theta}$ denotes the parameters of distribution.

In variational inference, the aim is to approximate the true posterior ${p}_{\tilde{\boldsymbol{y}}, \tilde{\boldsymbol{z}} \mid \boldsymbol{x}}$ with a parametric variational density $q_{\tilde{\boldsymbol{y}}, \tilde{\boldsymbol{z}} \mid \boldsymbol{x}}$. This is accomplished by minimizing the expected Kullback–Leibler (KL) divergence between $q_{\tilde{\boldsymbol{y}}, \tilde{\boldsymbol{z}} \mid \boldsymbol{x}}$ and ${p}_{\tilde{\boldsymbol{y}}, \tilde{\boldsymbol{z}} \mid \boldsymbol{x}}$ over the data distribution $p_{\bm{x}}$:
\begin{equation}
\begin{aligned}
& \min \mathbb{E}_{\boldsymbol{x} \sim {p}_{\boldsymbol{x}}} D_{\mathrm{KL}}\left[q_{\tilde{\boldsymbol{y}}, \tilde{\boldsymbol{z}} \mid \boldsymbol{x}}\| {p}_{\tilde{\boldsymbol{y}}, \tilde{\boldsymbol{z}} \mid \boldsymbol{x}}\right]= \\ & \min \mathbb{E}_{\boldsymbol{x} \sim p_{\boldsymbol{x}}} \mathbb{E}_{\tilde{\boldsymbol{y}}, \tilde{\boldsymbol{z}} \sim {q}} [\log {q}(\tilde{\boldsymbol{y}}, \tilde{\boldsymbol{z}} \mid \boldsymbol{x}) -\log {p}_{\boldsymbol{x} \mid \tilde{\boldsymbol{y}}}(\boldsymbol{x} \mid \tilde{\boldsymbol{y}}) \\ & -\log {p}_{\tilde{\boldsymbol{y}} \mid \tilde{\boldsymbol{z}}}(\tilde{\boldsymbol{y}} \mid \tilde{\boldsymbol{z}})-\log p_{\tilde{\boldsymbol{z}}}(\tilde{\boldsymbol{z}})]+\mathrm{const}.
\label{eq:KL divergence}
\end{aligned}
\end{equation}

Since $q_{\tilde{\boldsymbol{y}}, \tilde{\boldsymbol{z}} \mid \boldsymbol{x}}$ is composed of a product of uniform densities with unit width, the first item in (\ref{eq:KL divergence}) is zero and can be omitted:
\begin{equation}
\begin{aligned}
q(\tilde{\boldsymbol{y}}, \tilde{\boldsymbol{z}} \mid \boldsymbol{x}) =\prod_i \mathcal{U}(\tilde{y}_i; y_i, 1) \cdot \prod_j \mathcal{U}(\tilde{z}_j; z_j, 1),
\end{aligned}
\end{equation}
where $\mathcal{U}(\tilde{y}_i; y_i, 1)$ and $\mathcal{U}(\tilde{z}_j; z_j, 1)$ denote the uniform densities on the unit intervals centered at $y_i$ and $z_j$, respectively.

Minimizing the second item in (\ref{eq:KL divergence}) is equivalent to minimizing the expected distortion of the reconstructed image \cite{balle2018variational}, which can be understand as the distortion. The third and fourth item represent cross entropy as rates. Accordingly, the objective function can be further expressed as
\begin{equation}
\begin{aligned}
L = \lambda \cdot D + R & = \lambda \cdot \underbrace{\mathbb{E}_{\boldsymbol{x} \sim {p}_{\boldsymbol{x}}}\|\boldsymbol{x}-\hat{\boldsymbol{x}}\|_2^2}_{\text {distortion }}+\underbrace{\mathbb{E}_{\boldsymbol{x} \sim {p}_{\boldsymbol{x}}}\left[-\log _2 {p}_{\hat{\boldsymbol{y}}}(\hat{\boldsymbol{y}})\right]}_{\text {rate (latent features) }} \\ 
&+\underbrace{\mathbb{E}_{\boldsymbol{x} \sim {p}_{\boldsymbol{x}}}\left[-\log _2 {p}_{\hat{\boldsymbol{z}}}(\hat{\boldsymbol{z}})\right]}_{\text {rate (hyper latent features) }}.
\end{aligned}
\end{equation}

The core principle of compression lies in leveraging the sparsity of the input signal to eliminate redundancy and achieve a more compact representation. In essence, sparse signals are more likely to be compressed efficiently \cite{aberdam2019multi}. Thus, decomposing complex signals, such as SC images, into several frequency components where features vary smoothly offers a potential higher compression efficiency. In light of this, we propose a multi-frequency decomposition and adaptive quantization method for SC image compression. Therefore, the proposed objective function can be formulated as
\begin{equation}
\begin{aligned}
& L = \lambda \cdot D + R = \lambda \cdot \underbrace{\mathbb{E}_{\boldsymbol{x} \sim {p}_{\boldsymbol{x}}}\|\boldsymbol{x}-\hat{\boldsymbol{x}}\|_2^2}_{\text {distortion }} \\ & +\underbrace{\mathbb{E}_{\boldsymbol{x} \sim {p}_{\boldsymbol{x}}}-\left[\log _2 {p}_{\hat{\boldsymbol{y}}^H}(\hat{\boldsymbol{y}}^H)+\log _2 {p}_{\hat{\boldsymbol{y}}^M}(\hat{\boldsymbol{y}}^M)+\log _2 {p}_{\hat{\boldsymbol{y}}^L}(\hat{\boldsymbol{y}}^L)\right]}_{\text {rate (multi-frequency latent features) }} \\
&+\underbrace{\mathbb{E}_{\boldsymbol{x} \sim {p}_{\boldsymbol{x}}}-\left[\log _2 {p}_{\hat{\boldsymbol{z}}^H}(\hat{\boldsymbol{z}}^H)+\log _2 {p}_{\hat{\boldsymbol{z}}^M}(\hat{\boldsymbol{z}}^M)+\log _2 {p}_{\hat{\boldsymbol{z}}^L}(\hat{\boldsymbol{z}}^L)\right]}_{\text {rate (multi-frequency hyper latent features) }},
\end{aligned}
\label{eq:loss}
\end{equation}
where \(p_{\bm{x}}\) represents the true distribution of SC images, \({p}_{\bm{\hat{y}}^H \mid \bm{\hat{z}}^H}\), \({p}_{\bm{\hat{y}}^M \mid \bm{\hat{z}}^M}\), and \({p}_{\bm{\hat{y}}^L \mid \bm{\hat{z}}^L}\) correspond to the conditional entropy models, respectively,
\begin{flalign}
&\begin{aligned}
p_{\bm{\hat{y}}^H \mid \bm{\hat{z}}^H} & \left(\bm{\hat{y}}^H \mid \bm{\Phi}^H\right) = \\ & \hspace{-0.6cm} \prod_i\left(\mathcal{N}\left(\mu_i^H, \sigma_i^{H}\right) * \mathcal{U}\left(-\frac{\bm{\Delta}_H}{2}, \frac{\bm{\Delta}_H}{2}\right)\right)\left(\hat{y}_i^H\right),
\end{aligned}&\\
&\begin{aligned}
p_{\bm{\hat{y}}^M \mid \bm{\hat{z}}^M} & \left(\bm{\hat{y}}^M \mid \bm{\Phi}^M\right) = \\ & \hspace{-0.7cm} \prod_i\left(\mathcal{N}\left(\mu_i^M, \sigma_i^{M}\right) * \mathcal{U}\left(-\frac{\bm{\Delta}_M}{2}, \frac{\bm{\Delta}_M}{2}\right)\right)\left(\hat{y}_i^M\right),
\end{aligned}&\\
&\begin{aligned}
p_{\bm{\hat{y}}^L \mid \bm{\hat{z}}^L} & \left(\bm{\hat{y}}^L \mid \bm{\Phi}^L\right) = \\ & \hspace{-0.5cm} \prod_i\left(\mathcal{N}\left(\mu_i^L, \sigma_i^{L}\right) * \mathcal{U}\left(-\frac{\bm{\Delta}_L}{2}, \frac{\bm{\Delta}_L}{2}\right)\right)\left(\hat{y}_i^L\right),
\end{aligned}
\end{flalign}
where $\bm{\Delta}_H$,$\bm{\Delta}_M$ and $\bm{\Delta}_L$ represent noise interval parameters of different frequency components, and \(\bm{\hat{y}}^H\), \(\bm{\hat{y}}^M\), and \(\bm{\hat{y}}^L\) can be modeled as a convolution of a Gaussian distribution with a unit uniform distribution. The corresponding means (\(\mu_i^H\), \(\mu_i^M\), \(\mu_i^L\)) and scales (\(\sigma_i^{H}\), \(\sigma_i^{M}\), \(\sigma_i^{L}\)) are obtained through the entropy parameter networks \(P_h\), \(P_m\), and \(P_l\), respectively, where \(\bm{\Phi}^H\), \(\bm{\Phi}^M\), and \(\bm{\Phi}^L\) represent the learnable parameters of these networks.
\begin{figure*}
	\centering
	\begin{minipage}{\textwidth}
	\centering
	\includegraphics[width=\textwidth]{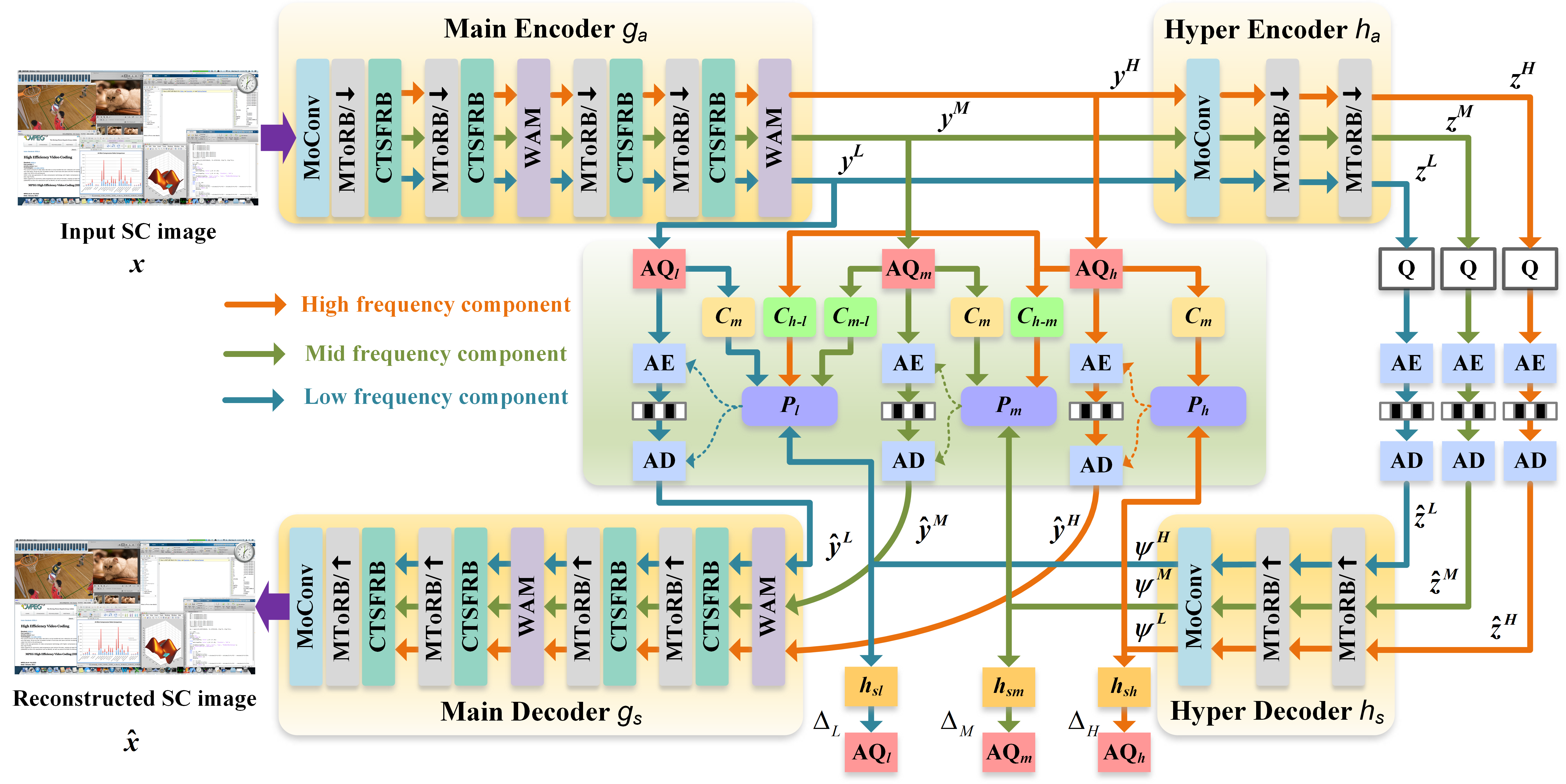}
	\caption{Framework of the proposed method. The codec consists of a main encoder-decoder and a hyper encoder-decoder. $\mathrm{MToRB}$ denotes multi-frequency two-stage octave residual block, $\mathrm{CTSFRB}$ denotes cascaded triple-scale feature fusion residual block, $C_m$ denotes context module, WAM denotes window-based attention block \cite{Zou_2022_CVPR}, $C_{h-l}$, $C_{h-m}$ and $C_{m-l}$ denote multi-frequency context interaction module (MFCIM). $P_h$, $P_m$ and $P_l$ denote entropy parameter network \cite{minnen2018joint} for $\bm{y}^H$,  $\bm{y}^M$ and $\bm{y}^L$. $\downarrow$ and $\uparrow$ denote downsampling and upsampling, respectively. The hyperprior $\bm{\mathit{\Psi}} ^H$, $\bm{\mathit{\Psi}} ^M$ and $\bm{\mathit{\Psi}} ^L$ are input into \(h_{sl}\), \(h_{sm}\), and \(h_{sh}\) to obtain the noise interval parameters $\bm{\Delta}_H$, $\bm{\Delta}_M$ and $\bm{\Delta}_L$, which are then applied to the adaptive quantization modules $\mathrm{AQ}_h$, $\mathrm{AQ}_m$ and $\mathrm{AQ}_l$, respectively.}
	\label{fig:framework}
	\end{minipage}
    \vspace{-0.3cm}
\end{figure*}

\section{Proposed method}

Building on the formulation of Section III, we propose an SC image compression method featuring a multi-frequency feature decomposition and adaptive quantization tailored for different frequency components, as depicted in Fig. \ref{fig:framework}. The framework is constructed based on the commonly used variational autoencoder (VAE). The base encoder \({g_a}\) decomposes the input feature \(\bm{x}\) into high-frequency latent features (\(\bm{y}^H\)), mid-frequency latent features (\(\bm{y}^M\)), and low-frequency latent features (\(\bm{y}^L\)). Subsequently, a hyper encoder \(h_a\) extracts the corresponding hyperprior latent features \(\bm{z}^H\), \(\bm{z}^M\), and \(\bm{z}^L\), which are utilized by the hyper decoder \(h_s\) to construct entropy models specific to each frequency component. Finally, the base decoder \(g_s\) obtains the reconstructed image \(\bm{\hat{x}}\) from latent features. 

Specifically, the proposed network includes four specified components, i.e., multi-frequency two-stage octave residual block, cascaded triple-scale feature fusion residual block, multi-frequency context interaction module and adaptive quantization module. 

\subsection{Multi-frequency two-stage octave residual block (MToRB)}
Compared to \cite{jiang2024OMR}, we further analyzed the spectral distribution of screen content images, identifying the non-smooth transition between extremely high and low frequencies and introduced a mid-frequency component and proposed MToRB, as illustrated in Fig. \ref{fig:octave methods} (d), to maintain high-frequency feature resolution while further reducing spatial redundancy in low-frequency features. The input features \(\bm{X}\) are divided along the channel dimension into high-frequency features \(\bm{X}^H\), mid-frequency features \(\bm{X}^M\), and low-frequency features \(\bm{X}^L\). Through the first-stage convolution operation (denoted as MoConv), we can obtain the features \(\bm{Y}^H_p\), \(\bm{Y}^M_p\), and \(\bm{Y}^L_p\):
\begin{flalign}
&\begin{aligned}
\bm{Y}^H_p & =\bm{Y}^{H \rightarrow H}+\bm{Y}^{M \rightarrow H}+\bm{Y}^{L \rightarrow H} \\
& =f\left(\bm{X}^H ; \bm{W}^{H \rightarrow H} ; 1\right)+f_{\uparrow}\left(\bm{X}^M ; \bm{W}^{M \rightarrow H} ; 2\right) \\ & +f_{\uparrow}\left(\bm{X}^L ; \bm{W}^{L \rightarrow H} ; 2\right),
\end{aligned}&\\
&\begin{aligned}
\bm{Y}^M_p & =\bm{Y}^{M \rightarrow M}+\bm{Y}^{H \rightarrow M}+\bm{Y}^{L \rightarrow M} \\ 
& =f\left(\bm{X}^M ; \bm{W}^{M \rightarrow M} ; 1\right)+f_{\downarrow}\left(\bm{X}^H ; \bm{W}^{H \rightarrow M} ; 2\right)\\
& +f_{\uparrow}\left(\bm{X}^L ; \bm{W}^{L \rightarrow M}; 2\right),
\end{aligned}&\\
&\begin{aligned}
\bm{Y}^L_p & =\bm{Y}^{L \rightarrow L}+\bm{Y}^{H \rightarrow L}+\bm{Y}^{M \rightarrow L}\\
& =f\left(\bm{X}^L ; \bm{W}^{L \rightarrow L} ; 1\right)+f_{\downarrow}\left(\bm{X}^H ; \bm{W}^{H \rightarrow L} ; 2\right)\\
& +f_{\downarrow}\left(\bm{X}^M ; \bm{W}^{M \rightarrow L} ; 2\right),
\end{aligned}
\end{flalign}
where the intra-frequency features include \(\bm{Y}^{H \rightarrow H}\), \(\bm{Y}^{M \rightarrow M}\), and \(\bm{Y}^{L \rightarrow L}\), while the inter-frequency features include \(\bm{Y}^{M \rightarrow H}\), \(\bm{Y}^{L \rightarrow H}\), \(\bm{Y}^{H \rightarrow M}\), \(\bm{Y}^{L \rightarrow M}\), \(\bm{Y}^{H \rightarrow L}\), and \(\bm{Y}^{M \rightarrow L}\). 
\begin{figure}[t]
	\centering
	\begin{minipage}{0.45\textwidth}
	\centering
	\includegraphics[width=\textwidth]{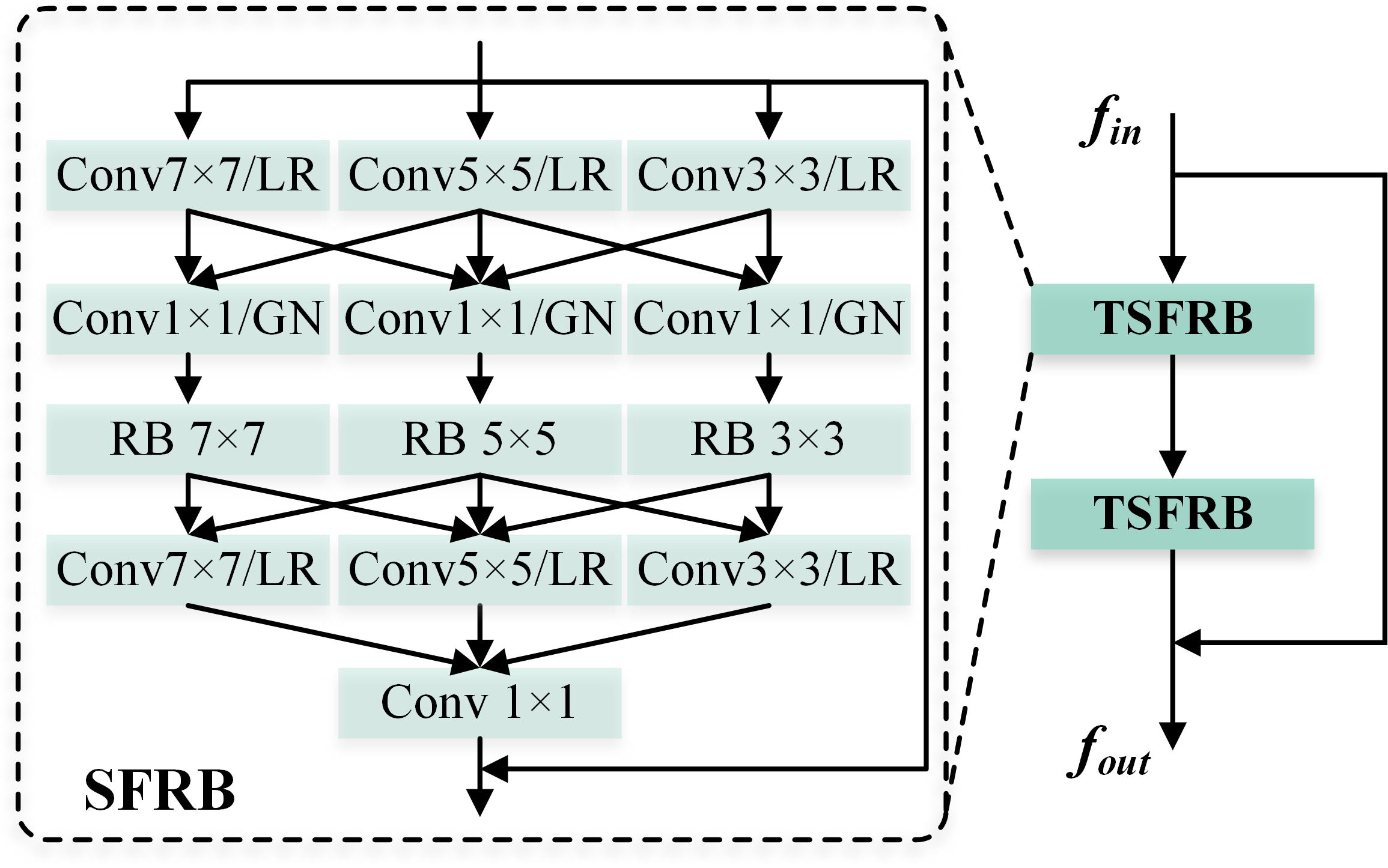}
	\caption{Structure of the CTSFRB, in which LR and GN represent Leaky ReLU and generalized divisive normalization \cite{balle2017end}.}
	\label{fig:ctsfrb}
	\end{minipage}
    \vspace{-0.3cm}
\end{figure}

The final output features \(\bm{Y}^H\), \(\bm{Y}^M\), and \(\bm{Y}^L\) are then produced through the second-stage downsampling convolutions and skip connections:
\begin{align}
\bm{Y}^H & =f_{\downarrow}\left(\bm{Y}^H_p ; \bm{W}^H ; 2\right)+f_{\mathrm{st}}\left(\bm{X}^H ; \bm{W}_{sc}^H ; 2\right), \\
\bm{Y}^M & =f_{\downarrow}\left(\bm{Y}^M_p ; \bm{W}^M ; 2\right)+f_{\mathrm{st}}\left(\bm{X}^M ; \bm{W}_{sc}^M ; 2\right), \\
\bm{Y}^L & =f_{\downarrow}\left(\bm{Y}_p^L ; \bm{W}^L ; 2\right)+f_{\mathrm{st}}\left(\bm{X}^L ; \bm{W}_{sc}^L ; 2\right).
\end{align}

\subsection{Cascaded triple-scale feature fusion residual block (CTSFRB)}
The CTSFRB is proposed to handle the multi-frequency features of SC image. Unlike existing multi-scale residual blocks \cite{li2018multi}, our approach utilizes convolutional kernels of three different scales to extract features at varying resolutions, enabling feature interaction across these scales. The cascaded structure further enhances the interaction between different scale features and improves the network's nonlinear representation capabilities. As shown in Fig. \ref{fig:ctsfrb}, each CTSFRB is composed of two triple-scale feature fusion residual blocks (TSFRBs) connected via skip connections.

In each TSFRB, different branches use convolutional kernels of varying sizes to extract features across multiple scales. Specifically, for the input feature \(\bm{f}_{in}\), the first branch employs a convolutional layer with a 3×3 kernel, the second branch uses a 5×5 kernel, and the third branch utilizes a 7×7 kernel. The features extracted from these three branches are concatenated for an initial feature fusion. After passing through a 1×1 convolution, the features are sent to their corresponding residual blocks and convolutional layers for further processing. A second feature fusion is then performed, where the three resulting features are concatenated and fed into a 1×1 convolutional layer. The output is then combined with a skip connection from the original input feature to produce the final output feature \(\bm{f}_{out}\) of the current TSFRB.
\subsection{Multi-frequency context interaction module (MFCIM)}
In the proposed MToRB, the single latent feature is decomposed into three distinct frequency components and there are also residual correlations between these frequency components, which can not be fully removed. Existing context networks typically target feature components of the same frequency, failing to capture the correlations between different frequency components. The work in \cite{chen2022two} demonstrated the effectiveness of information transfer between high- and low-frequency features. To further eliminate correlations between frequency components, we propose an MFCIM which constructs context interaction networks between high-, mid-, and low-frequency components.

\begin{figure}[t]
\begin{minipage}[t]{\linewidth}
  \centering
  \includegraphics[width=0.6\linewidth]{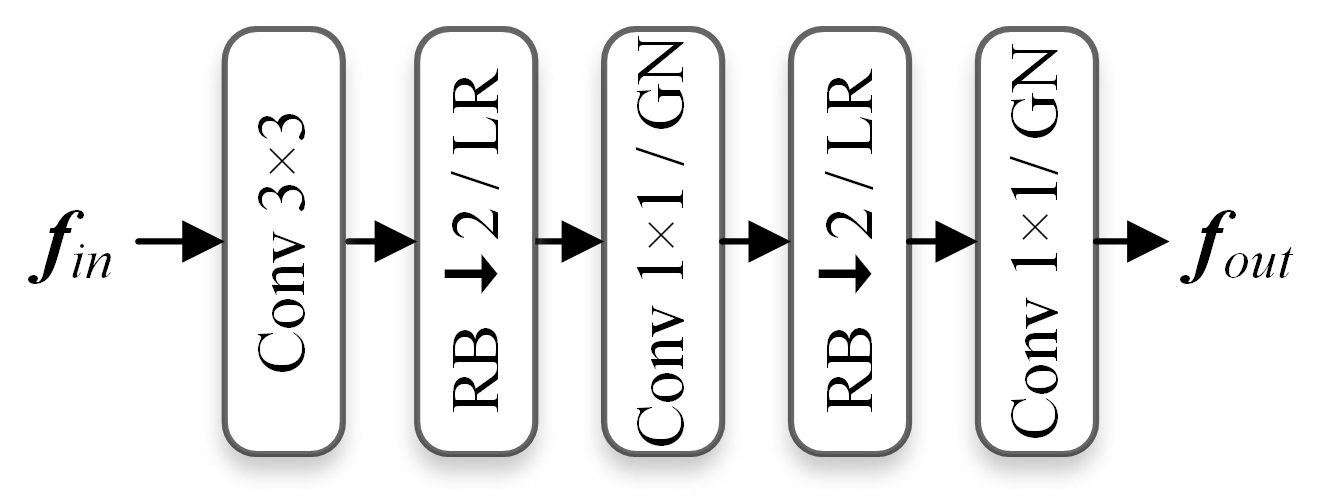}
  \centerline{(a)}\medskip
\end{minipage}
\begin{minipage}[t]{\linewidth}
  \centering
  \includegraphics[width=0.6\linewidth]{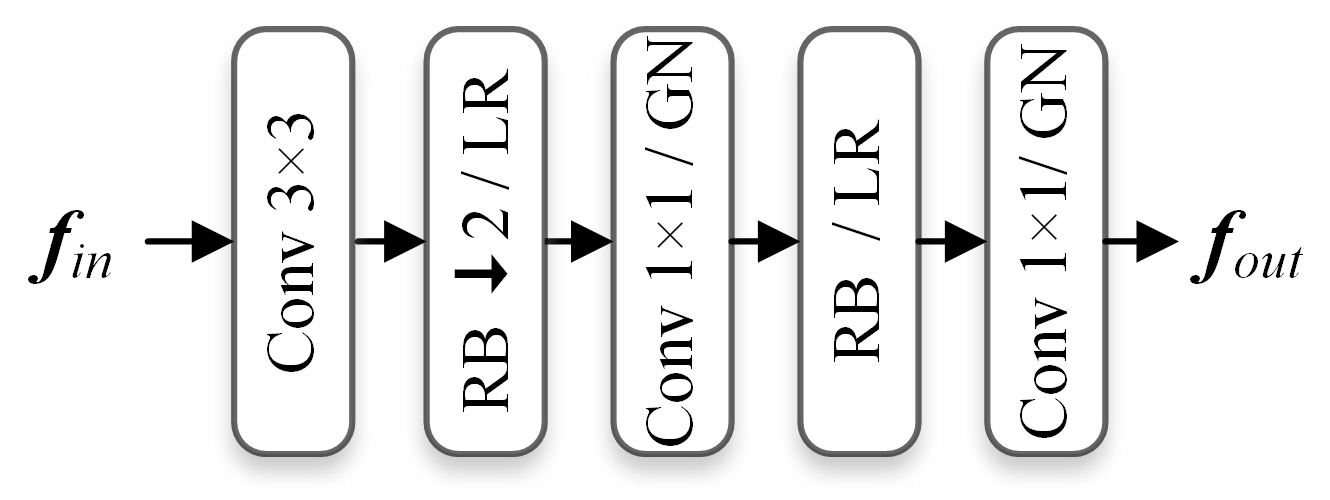}
  \centerline{(b)}\medskip
\end{minipage}
\caption{Multi-frequency context interaction module (MFCIM), (a) the structure of $C_{h-l}$, (b) the structure of $C_{h-m}$ and $C_{m-l}$.}
\label{fig:context interaction}
\vspace{-0.2cm}
\end{figure}

Following the structure in Fig. \ref{fig:LIC framework}, we adopted a multi-stage context model \cite{lu2022high} as $C_m$. The context interaction module from high to low frequency is denoted as $C_{h-l}$, from high to mid frequency as $C_{h-m}$, and from mid to low frequency as $C_{m-l}$. As illustrated in Fig. \ref{fig:context interaction}, $C_{h-l}$ utilizes two downsampling residual block (RB) convolutions with a stride of 2, while $C_{h-m}$ and $C_{m-l}$ use only one downsampling RB convolution with a stride of 2. To efficiently estimate the entropy model parameters of the low-frequency latent feature $\bm{\hat{y}}^L$, we not only use the low-frequency context module $C_m$, but also introduce $C_{h-l}$ from the high-frequency component and $C_{m-l}$ from the mid-frequency component. As shown in Fig. \ref{fig:framework}, the outputs of $C_m$, $C_{h-l}$, and $C_{m-l}$, along with the low-frequency hyperprior 
$\bm{\mathit{\Psi}} ^L$, are concatenated and fed into the entropy parameter network $P_l$ to obtain the distribution parameters of $\bm{\hat{y}}^L$.
\begin{figure}[t]
	\centering
	\begin{minipage}{0.45\textwidth}
	\centering
	\includegraphics[width=0.7\textwidth]{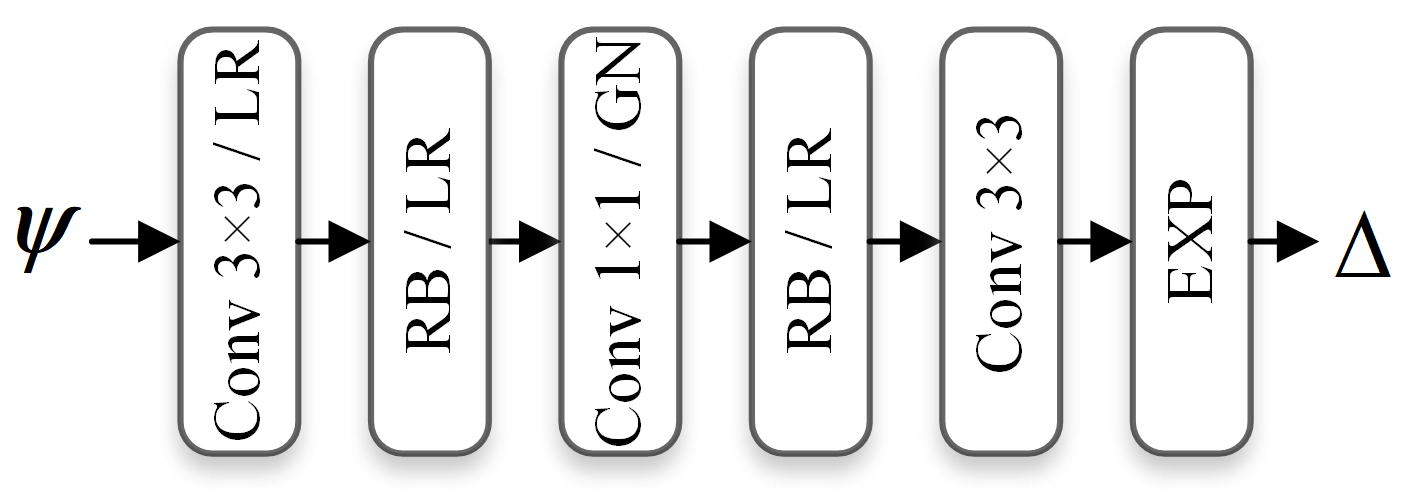}
	\caption{Structure of the quantization interval modules.}
	\label{fig:quantization}
	\end{minipage}
    \vspace{-0.2cm}
\end{figure}
\subsection{Adaptive quantization (AQ) module}
Existing LIC methods use additive uniform noise during training to replace the rounding operation in the testing phase. Although this approach approximates quantization error with additive uniform noise, it introduces a mismatch between training and testing. Additionally, because the noise interval parameter is fixed, it prevents adaption of the quantization step size during testing. To address this issue, Guo \textit{et al}. \cite{guo2021soft} proposed a network module for learning scaled uniform noise. However, this method applies the same quantization step size to different features, disregarding their individual discrepancy, which limits the performance. Therefore, we propose separate network branches to learn the quantization interval parameters for different frequency components, allowing for more flexible control of quantization granularity across different frequency features, as illustrated in Fig. \ref{fig:framework}. The high-frequency noise interval parameter $\bm{\Delta}_H$, mid-frequency noise interval parameter $\bm{\Delta}_M$ and low-frequency noise interval parameter $\bm{\Delta}_L$ can be determined from the quantization-interval modules $h_{sh}$, $h_{sm}$ and $h_{sl}$, as shown in Fig. \ref{fig:quantization}. 

Taking the high-frequency component as an example, \(h_s\) generates the high-frequency features hyperprior \(\bm{\mathit{\Psi}} ^H\) from $\bm{\tilde{z}}^H$, and then \(h_{sh}\) generates the high-frequency noise interval parameter $\bm{\Delta}_H$ from \(\bm{\mathit{\Psi}} ^H\): 
\begin{align}
\bm{\mathit{\Psi}} ^H &=h_s\left(\bm{\tilde{z}}^H\right),\\
\bm{\Delta}_H &=h_{s h}\left(\bm{\mathit{\Psi}}^H\right).
\end{align}

During the training stage, the quantized high-frequency latent feature $\bm{\tilde{y}}^H$ is obtained by adding the random scaled uniform noise \(\bm{u}^H\) in the interval $[\bm{-\Delta}_H / 2,\bm{\Delta}_H / 2]$ to \(\bm{y}^H\): 
\begin{equation}
\bm{\tilde{y}}^H=\bm{y}^H+\bm{u}^H, \bm{u}^H \sim U\left(-\frac{\bm{\Delta}_H}{2}, \frac{\bm{\Delta}_H}{2}\right).
\end{equation}

Therefore, the variational density function $q(\tilde{\boldsymbol{y}}^H, \tilde{\boldsymbol{z}}^H \mid \boldsymbol{x}^H)$ in the first term of (\ref{eq:KL divergence}) depends on \(\bm{\Delta}_H\):
\begin{equation}
\begin{aligned}
q(\tilde{\boldsymbol{y}}^H, \tilde{\boldsymbol{z}}^H \mid \boldsymbol{x}^H) &=\prod_i \mathcal{U}(\tilde{y}_i^H; y_i^H, \bm{\Delta}_H) \cdot \prod_j \mathcal{U}(\tilde{z}_j^H; z_j^H, 1) \\ 
& =\frac{1}{\bm{\Delta}_H},
\end{aligned}
\end{equation}
\begin{figure*}[t]
	\centering
	\begin{minipage}{\textwidth}
	\centering
	\includegraphics[width=\textwidth]{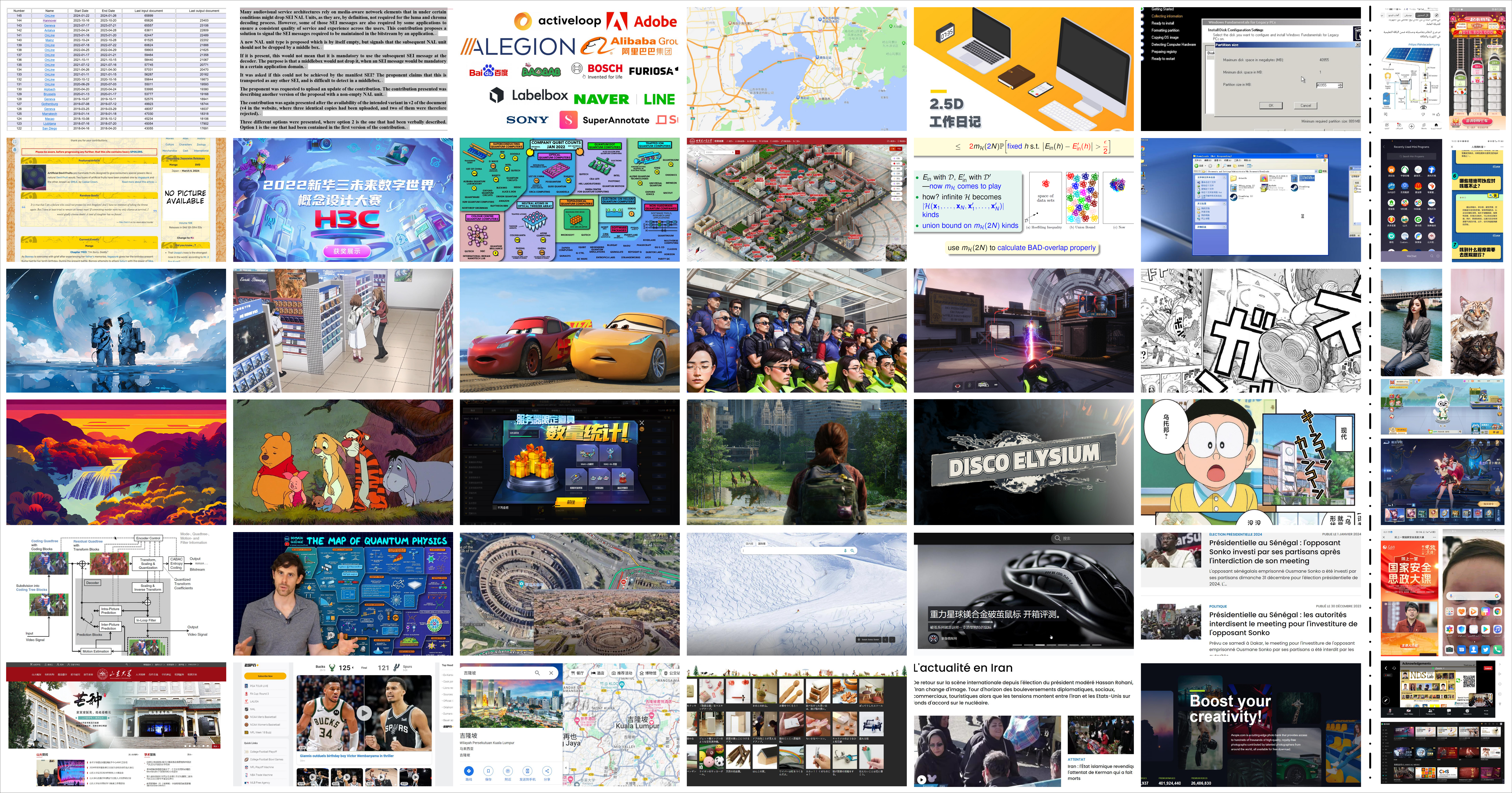}
	\caption{Example images from SDU-SCICD10K. The left side displays images from the PC platform, while the right side shows images from the mobile platform. The first two rows represent web and office SC images, the middle two rows depict computer-rendered images, and the bottom two rows illustrate mixed NS and SC images.}
	\label{fig:dataset}
	\end{minipage}
\end{figure*}

Accordingly, (\ref{eq:density}) can be can be expressed as
\begin{equation}
p_{\bm{\tilde{y}}^{H} \mid \bm{\tilde{z}}^H}\left(\bm{\tilde{y}}^{H} \mid \bm{\tilde{z}}^H\right)=\int_{\bm{\tilde{y}}^{H}-\frac{\bm{\Delta}_H}{2}}^{\bm{\tilde{y}}^{H}+\frac{\bm{\Delta}_H}{2}} \frac{1}{\bm{\Delta}_H} {p}\left(\bm{y}^{H}\right) d \bm{y}^{H},
\end{equation}
where $p_{\bm{\tilde{y}}^{H} \mid \bm{\tilde{z}}^H}\left(\bm{\tilde{y}}^{H} \mid \bm{\tilde{z}}^H\right)$ denotes the density distribution of entropy model. 

During the testing stage, quantization is performed using the rounding operation, with the quantization step size determined by the noise interval parameter,
\begin{align}
\bm{\Delta}_H &=h_{s h}\left(h_s\left(\bm{\hat{z}}^H\right)\right),\\
\bm{\hat{y}}^H &=\bm{\Delta}_H \cdot\left\lceil\frac{\bm{y}^H}{\bm{\Delta}_H}\right\rfloor,
\end{align}
where $\bm{\hat{y}}^H$ represents the quantized high-frequency latent feature during testing, obtained by rounding the quotient between \(\bm{y}^H\) and the high-frequency quantization step size.

We can readily extend the approach from high-frequency to mid-frequency and low-frequency components, i.e., the noise interval parameter $\bm{\Delta}_M$ for mid-frequency components and $\bm{\Delta}_L$ for low-frequency components. They can be obtained by $h_{sm}$ and $h_{sl}$:
\begin{align}
\bm{\Delta}_M &= h_{s m}\left(h_s\left(\bm{\hat{z}}^M\right)\right), \\
\bm{\Delta}_L &= h_{s l}\left(h_s\left(\bm{\hat{z}}^L\right)\right), \\
\bm{\hat{y}}^M &= \bm{\Delta}_M \cdot\left\lceil\frac{\bm{y}^M}{\bm{\Delta}_M}\right\rfloor, \\
\bm{\hat{y}}^L &= \bm{\Delta}_L \cdot\left\lceil\frac{\bm{y}^L}{\bm{\Delta}_L}\right\rfloor,
\end{align}
where $\bm{\hat{y}}^M$ and $\bm{\hat{y}}^L$ represent the quantized mid-frequency and low-frequency features during testing, respectively.

\subsection{Loss function}
As shown in (\ref{eq:loss}), the objective of the proposed method is to jointly optimize the RD trade-off: 
\begin{equation}
\begin{aligned}
L & =\lambda \cdot D(\bm{x}, \bm{\hat{x}})+R_{\bm{y}}+R_{\bm{z}}\\
& =\lambda \cdot D(\bm{x}, \bm{\hat{x}})+R_{\bm{y}^H}+R_{\bm{y}^M}+R_{\bm{y}^L} \\ & +R_{\bm{z}^H}+R_{\bm{z}^M}+R_{\bm{z}^L},
\end{aligned}
\end{equation}
where \(D\) represents the distortion which can be measured using metrics such as mean square error (MSE) or multi-scale structural similarity (MS-SSIM). The Lagrange multiplier $\lambda$ controls the balance between bitrate and distortion. The bitrate \(R\) is composed of three parts $R^H$, $R^M$ and $R^L$, corresponding to the bitrates of the high-, mid-, and low-frequency components, respectively. Each component's rate is determined by the corresponding latent features $\bm{y}$ and the hyper latent features $\bm{z}$:
\begin{flalign}
&\begin{aligned}
R^H &= R_{\bm{y}^H}+ R_{\bm{z}^H} \\ 
    &= \mathbb{E}_{\bm{x} \sim p_{\bm{x}}}\left[-\log_2 {p}_{\bm{\hat{y}}^H \mid \bm{\hat{z}}^H}\left(\bm{\hat{y}}^H \mid \bm{\hat{z}}^H\right)\right] \\
    &\quad +\mathbb{E}_{\bm{x} \sim p_{\bm{x}}}\left[-\log_2 {p}_{\bm{\hat{z}}^H \mid \bm{\Theta}^H}\left(\bm{\hat{z}}^H \mid \bm{\Theta}^H\right)\right],
\end{aligned}&\\
&\begin{aligned}
R^M &= R_{\bm{y}^M}+ R_{\bm{z}^M} \\ 
    &= \mathbb{E}_{\bm{x} \sim p_{\bm{x}}}\left[-\log_2 {p}_{\bm{\hat{y}}^M \mid \bm{\hat{z}}^M}\left(\bm{\hat{y}}^M \mid \bm{\hat{z}}^M\right)\right] \\
    &\quad +\mathbb{E}_{\bm{x} \sim p_{\bm{x}}}\left[-\log_2 {p}_{\bm{\hat{z}}^M \mid \bm{\Theta}^M}\left(\bm{\hat{z}}^M \mid \bm{\Theta}^M\right)\right],
\end{aligned}&\\
&\begin{aligned}
R^L &= R_{\bm{y}^L}+ R_{\bm{z}^L} \\ 
    &= \mathbb{E}_{\bm{x} \sim p_{\bm{x}}}\left[-\log_2 {p}_{\bm{\hat{y}}^L \mid \bm{\hat{z}}^L}\left(\bm{\hat{y}}^L \mid \bm{\hat{z}}^L\right)\right] \\
    &\quad +\mathbb{E}_{\bm{x} \sim p_{\bm{x}}}\left[-\log_2 {p}_{\bm{\hat{z}}^L \mid \bm{\Theta}^L}\left(\bm{\hat{z}}^L \mid \bm{\Theta}^L\right)\right],
\end{aligned}
\end{flalign}
where $\bm{\Theta}^H$, $\bm{\Theta}^H$ and $\bm{\Theta}^H$ denote the distribution parameters of $\bm{z}^H$, $\bm{z}^M$ and $\bm{z}^L$, respectively, as illustrated in (\ref{eq:factorized}).

\begin{figure*}
	\centering
	\begin{minipage}[b]{0.47\linewidth}
	  \includegraphics[width=0.9\linewidth]{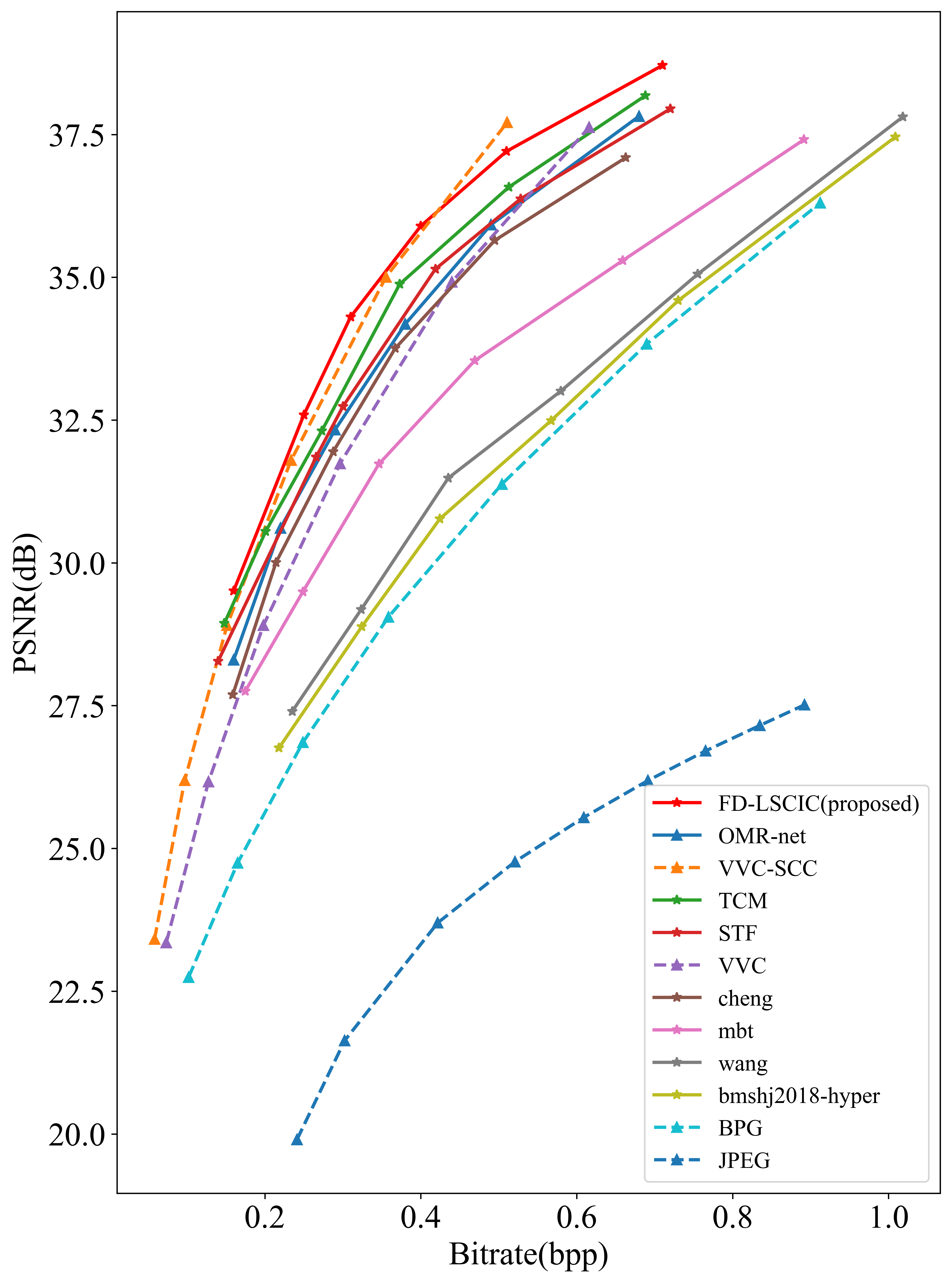}
	  \centerline{(a) SCID}\medskip
	\end{minipage}
	\begin{minipage}[b]{0.48\linewidth}
	  \includegraphics[width=0.9\linewidth]{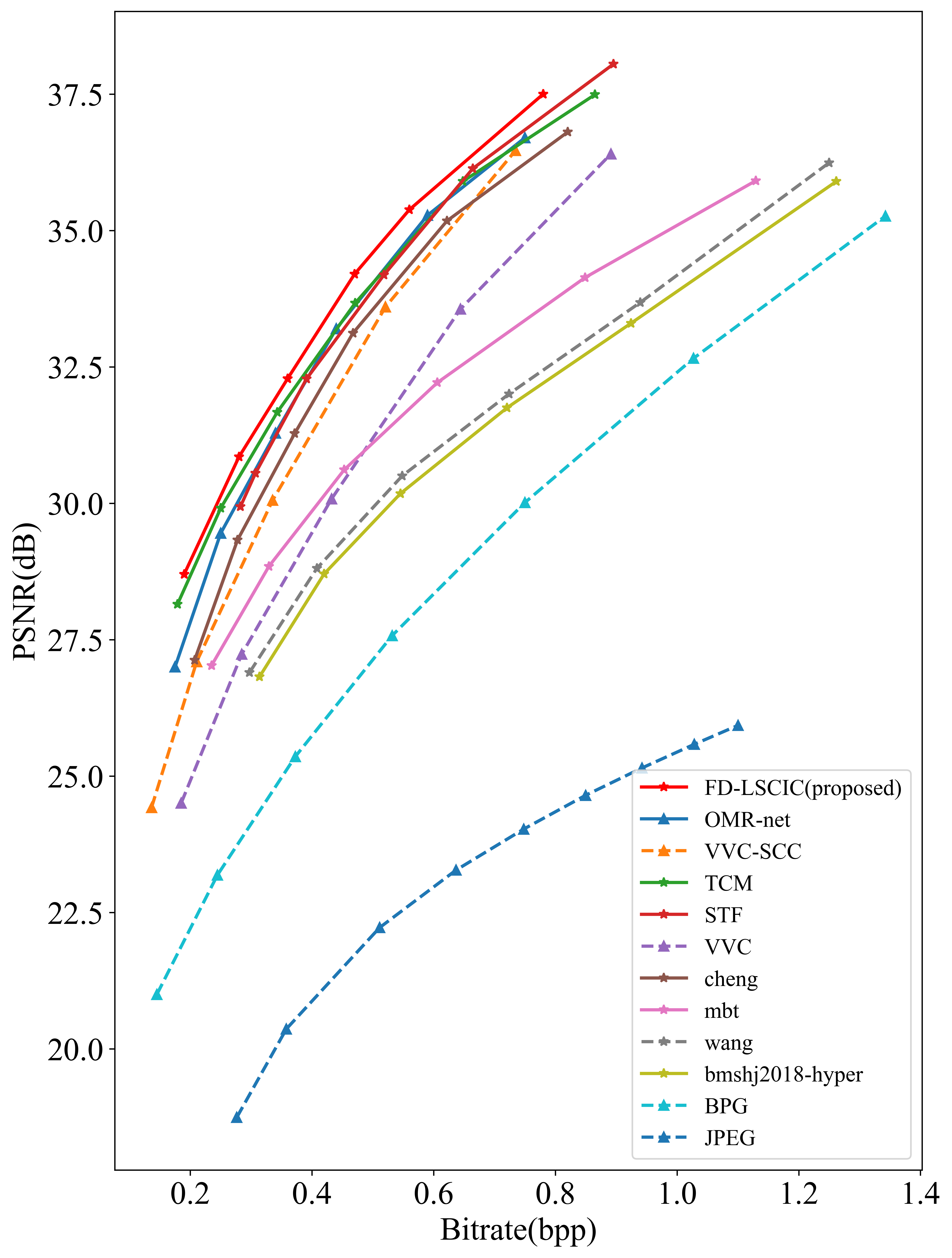}
	  \centerline{(b) SIQAD}\medskip
	\end{minipage}
	\centering
	\caption{Rate-PSNR curves comparison on the SCID and SIQAD datasets. The bitrate is expressed in bits per pixels (bpp).}
	\label{fig:mse_rd_curve}
    \vspace{-0.4cm}
\end{figure*}
\section{Experimental results and analysis}
\subsection{Screen content image compression dataset}
Existing SC datasets \cite{ni2017esim, yang2015perceptual, min2017unified} primarily focus on quality assessment \cite{Li2021subjective,Kwong2024deeplearning} and include only a small number of reference images. Publicly available datasets specifically designed for SC image compression are rare. In our previous work, we proposed a small screen content image dataset (SDU-SCICD2K) \cite{jiang2024OMR} containing over 2,000 images, which includes various types of content such as text, charts, graphics, animations, games, and mixed NS and SC images.

Building on this, we expanded the dataset to include a wider variety and greater quantity of images, and extended it by adding SC images coming from mobile platforms. Accordingly, we propose a large screen content image compression dataset (SDU-SCICD10K) comprising 10,000 images, which is broadly categorized into three types, each containing images from both PC and mobile platforms:

\textbf{Web and office SC images.} This category includes text, charts, simple graphics, webpages, PPTs, maps, and computer application interfaces. It covers SC generated by computers in daily use, often featuring simple and repetitive backgrounds. The text category encompasses multiple languages, including English, Chinese, Japanese, French, Latin, Russian, and Spanish, with resolutions ranging from 720p to 2K.

\textbf{Computer-rendered images.} This category includes animations, games, comics, and AI-generated images. These scenes cater to human visual needs and often feature more complex and dynamic backgrounds and content, with resolutions ranging from 720p to 4K.

\textbf{Mixed NS and SC images.} This category represents the common scenario where NS are partially or entirely embedded into SC, making them distinct yet interrelated. The resolutions in this category range from 720p to 2K.
\begin{figure*}
	\centering
	\begin{minipage}[b]{0.47\linewidth}
	  \includegraphics[width=0.9\linewidth]{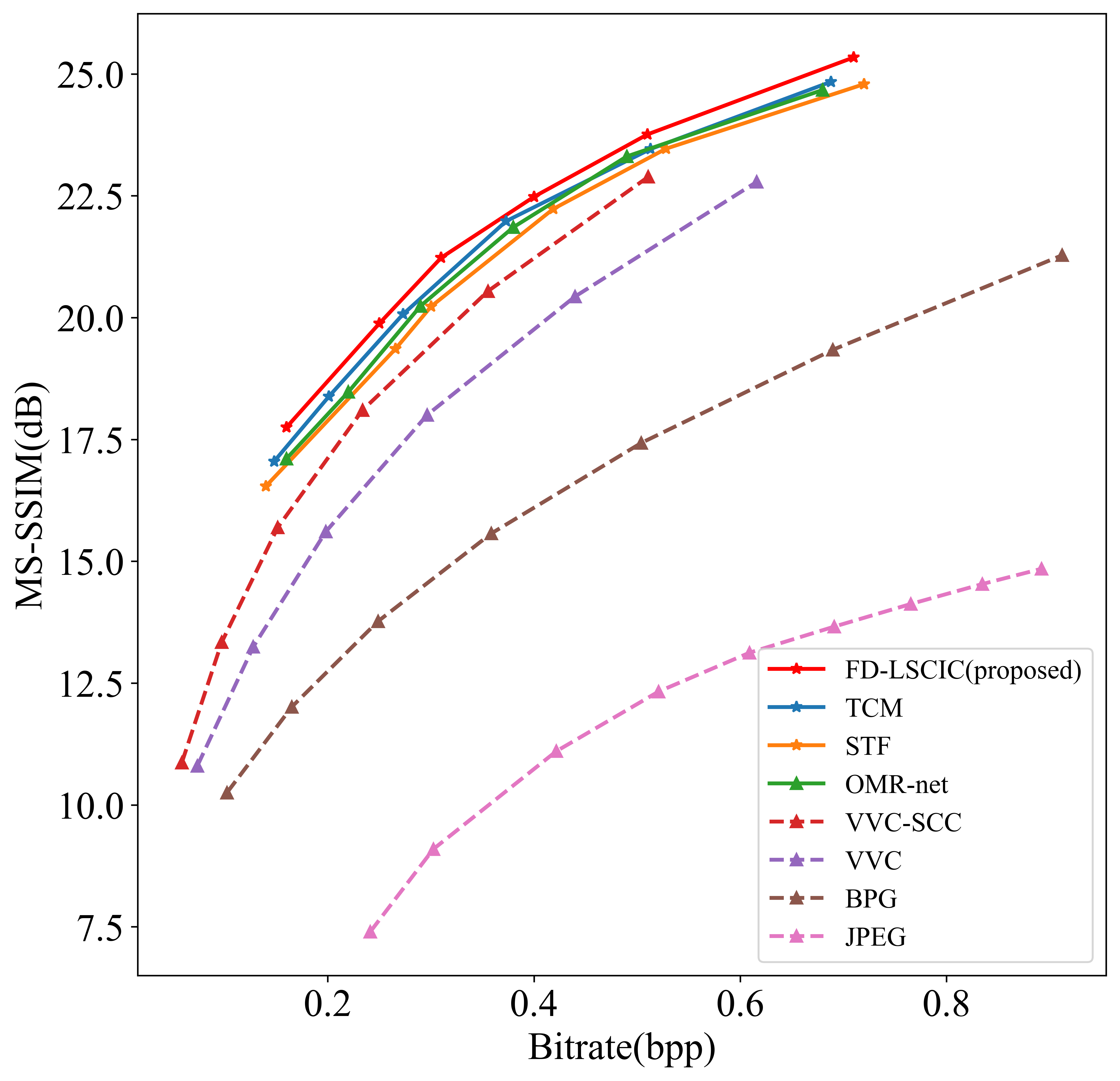}
	  \centerline{(a) SCID}\medskip
	\end{minipage}
	\begin{minipage}[b]{0.48\linewidth}
	  \includegraphics[width=0.9\linewidth]{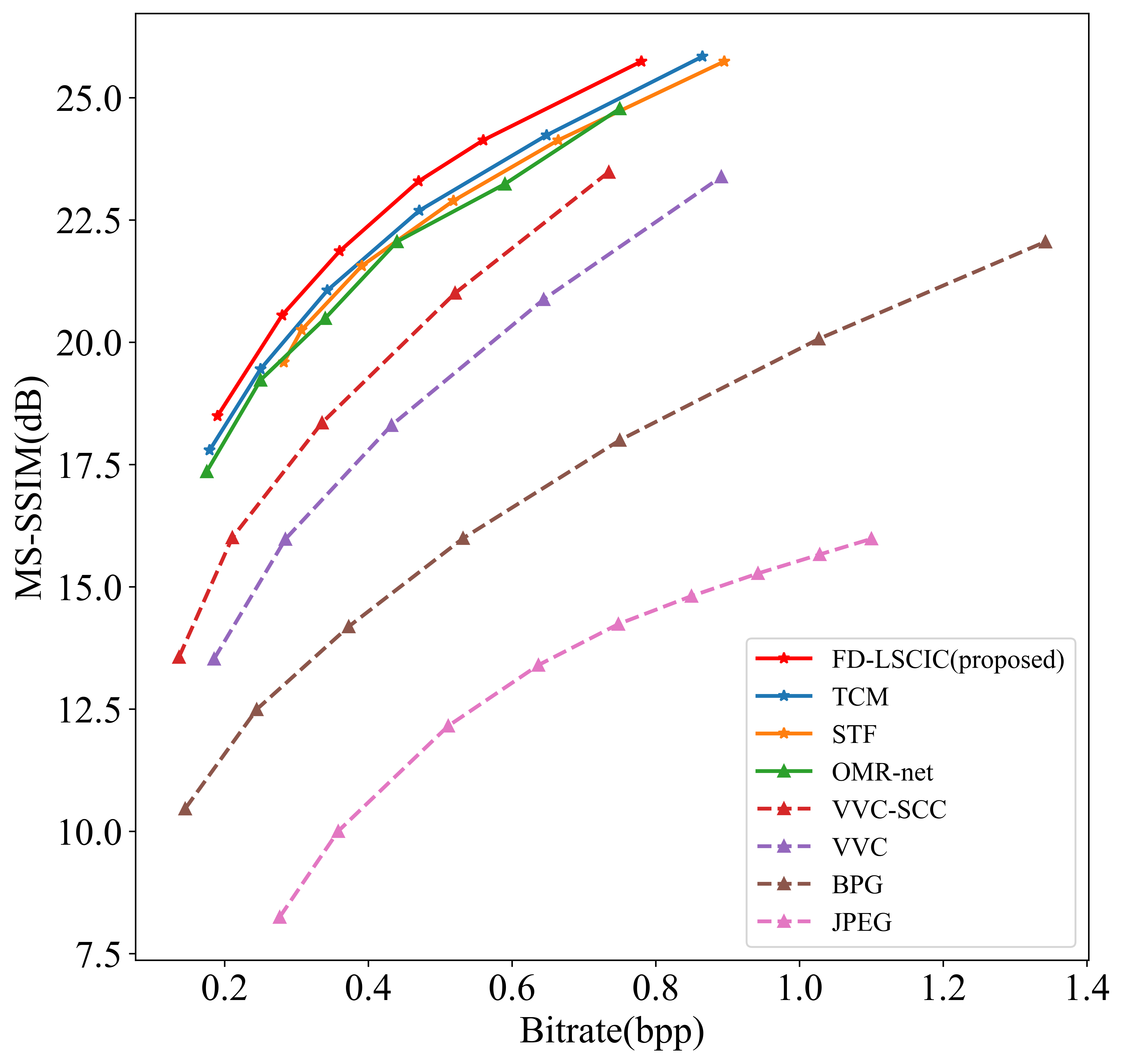}
	  \centerline{(b) SIQAD}\medskip
	\end{minipage}
	\centering
	\caption{Rate-MS-SSIM curves comparison on the SCID and SIQAD datasets. The bitrate is expressed in bits per pixels (bpp).}
	\label{fig:msssim_rd_curve}
\end{figure*}
\begin{table*}
  \begin{center}
	  \caption{Average BD-rate(\%) improvement and complexity of different methods on the SCID and SIQAD datasets using GPU (RTX4090). H.266/VVC (main profile without SCC techniques) is used as the anchor. Bold font indicates the best performance.}
	  \label{tab:bd-rate}
	  \begin{tabular}{c|cc|cc|c}
	    \toprule
        \multirow{2}{*}{\textbf{Methods}}& \multicolumn{2}{c|}{\textbf{BD-rate(\%) $\downarrow$}} & \multicolumn{2}{c|}{\textbf{Time complexity}} & \multicolumn{1}{c}{\textbf{Computational complexity}}\\
        & \textbf{SCID} & \textbf{SIQAD} & \textbf{Enc.Time (s)} & \textbf{Dec.Time (s)} & \textbf{Parameters (M)}\\
	    \midrule
		H.266/VVC \cite{pfaff2021intra} & 0.0  & 0.0 & 67.14 & 0.22 & / \\
        JPEG \cite{wallace1991jpeg} &  432.8  & 332.3 & 0.02 & 0.01 & /\\
        BPG \cite{bellard2017bpg} &  74.4  & 73.3 & 0.48 & 0.18 & / \\
        bmshj2018-hyper \cite{balle2018variational} &  67.3  & 33.4 & 2.74 & 4.12 & 11.8 \\
		wang \cite{wang2022transform} &  60.3  & 28.2 & 2.89  & 4.25 & 12.3 \\
        mbt \cite{minnen2018joint} &  24.8  & 5.1 & 28.73 & 35.29 & 25.5 \\
		cheng \cite{cheng2020learned} &  -2.8  & -24.3 & 12.16 & 25.78 & 29.6 \\
		STF \cite{Zou_2022_CVPR} & -9.6  & -28.0 & 0.98 & 1.22 & 99.9\\
		TCM \cite{liu2023learned}  &  -14.6  & -33.2 & 0.47 & 0.52  & 44.9 \\
        H.266/VVC-SCC \cite{bross2021overview} & -21.9  & -22.3 & 126.71 & 0.23  & / \\
        OMR-net \cite{jiang2024OMR} & -7.1  & -30.9 & 18.35  & 36.55 & 30.6\\ 
		\textbf{FD-LSCIC(proposed)}  &  \textbf{-22.4}  & \textbf{-36.4} & 1.2 & 1.4 & 59.5 \\
	  \bottomrule
	\end{tabular}
  \end{center}
\end{table*}
\subsection{Experimental Setup}
\textbf{Training.} We used the proposed SDU-SCICD10K dataset as the training set and randomly cropped all images into patches with size of $256 \times 256$. The number of epochs was set to 400. The learning rate was set to 1e-4 in the first 300 epochs, to 1e-5 in the following 100 epochs. We set $\alpha = \beta = \frac{1}{3}$, apart from the first and last MoConv where $\alpha_{in} = \beta_{in} = 0$, $\alpha_{out} = \alpha $, $\beta_{out} = \beta$ and $\alpha_{out} = \beta_{out} = 0$, $\alpha_{in} = \alpha $, $\beta_{in} = \beta$, respectively. The number of channels was set to 192 for lower-rate model and 320 for higher-rate model. 

\textbf{Settings.} We implemented the proposed method with the compressAI Pytorch library \cite{begaint2020compressai}, using the Adam optimizer with a batch size of 8 on Nvidia RTX4090. We set $\lambda$ to 0.0018, 0.0035, 0.0067, 0.013, 0.025, 0.0483 to minimize MSE. 

\textbf{Testing.} We evaluated the proposed method on two different SC image datasets:

SCID \cite{ni2017esim}. It contains 40 images with a resolution of $1280 \times 720$, covering content such as text, graphics, symbols, patterns, and some natural images, sourced from webpages, slides, comics, and digital magazines.

SIQAD \cite{yang2015perceptual}. It contains 20 images with varying resolutions around $600 \times 900$, covering content such as text, graphics, patterns, and some natural images, sourced from webpages, slides, and digital magazines.

\begin{figure*}
	\centering
	\begin{minipage}{\textwidth}
	\centering
	\includegraphics[width=0.9\textwidth]{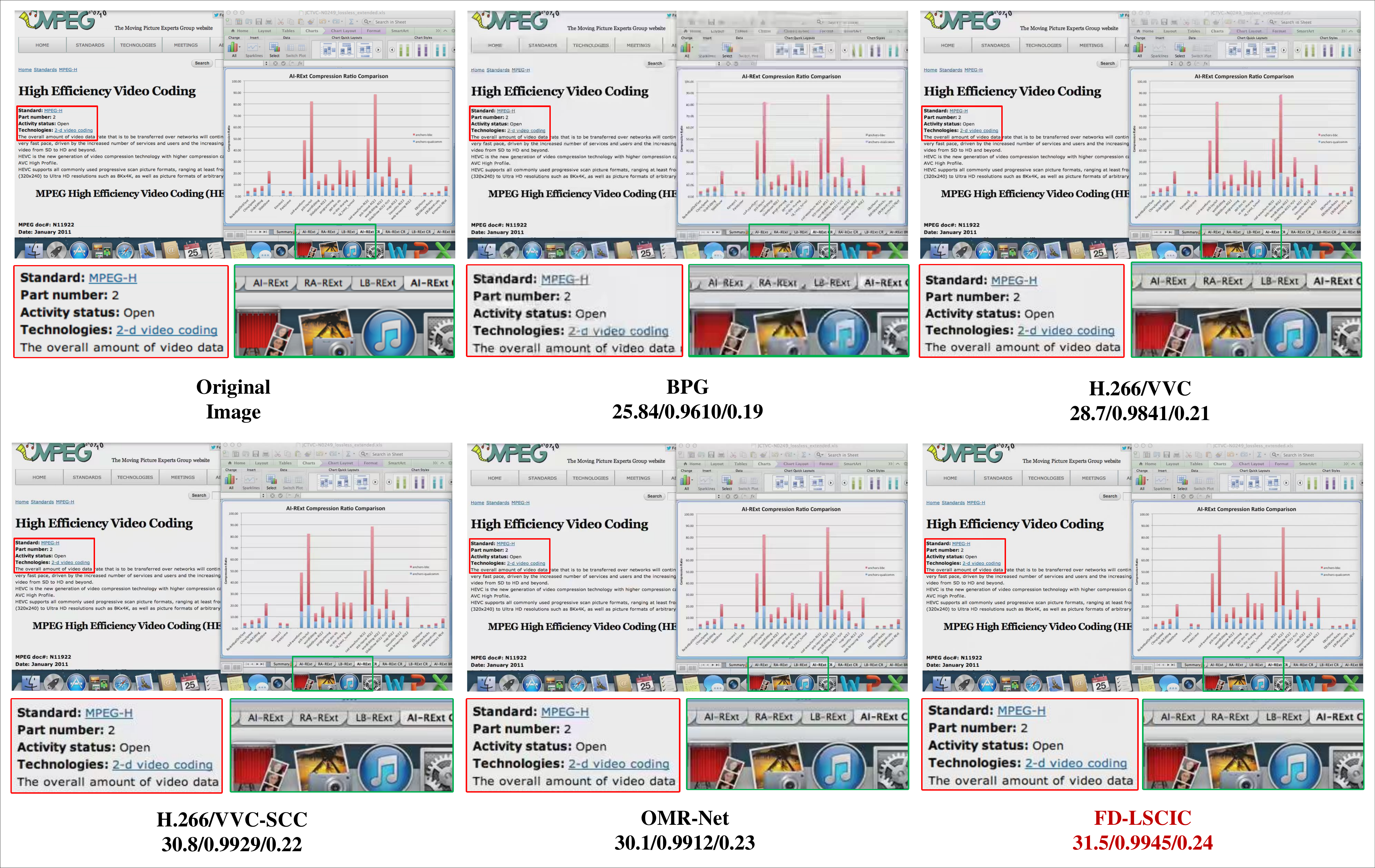}
	\caption{Visualization of “SCI36.png” in SCID dataset. For each method, we give the PSNR(dB)/MS-SSIM/bitrate(bpp).}
	\label{fig:subject}
	\end{minipage}
\end{figure*}

\subsection{RD Performance}
To evaluate the effectiveness of the proposed method (namely FD-LSCIC), we compared it with some well-known image compression standards, including JPEG \cite{wallace1991jpeg}, BPG \cite{bellard2017bpg}, H.266/VVC \cite{pfaff2021intra} and H.266/VVC-SCC \cite{bross2021overview} (currently the best codec for SC images), and the recent LIC methods, including bmshj2018-hyper \cite{balle2018variational}, mbt \cite{minnen2018joint}, cheng \cite{cheng2020learned}, wang \cite{wang2022transform}, STF \cite{Zou_2022_CVPR}, TCM \cite{liu2023learned} and OMR-net \cite{jiang2024OMR}. All models were retrained using the SDU-SCICD10K dataset, and the testing code was implemented based on the CompressAI platform \cite{begaint2020compressai}. Fig. \ref{fig:mse_rd_curve} and Fig. \ref{fig:msssim_rd_curve}, respectively, show the Rate-PSNR and Rate-MS-SSIM curves of the methods on the SCID and the SIQAD datasets. We can see that the proposed method achieves the best performance in terms of both PSNR and MS-SSIM. Table \ref{tab:bd-rate}, in which H.266/VVC is used as the anchor, shows the average Bjontegaard delta (BD-rate) \cite{bjontegaard2001calculation} reductions on the test sets. When tested on the SCID dataset, the proposed FD-LSCIC achieves a 22.4\% BD-rate reduction compared to H.266/VVC, while H.266/VVC-SCC achieves 21.9\% BD-rate reduction. When tested on the SIQAD dataset, the proposed FD-LSCIC reduces BD-rate by 36.4\%, while H.266/VVC-SCC achieves only 22.3\% BD-rate reduction. Compared to state-of-the-art methods based on LIC, our method achieves superior performance. Specifically, compare to TCM, FD-LSCIC reduces the BD-rate by 8\% on the SCID dataset and by 3.2\% on the SIQAD dataset. Compared to OMR-net, FD-LSCIC reduces the BD-rate by 15.3\% on SCID and by 5.5\% on SIQAD. In addition, more results indicate that the performance improvement of LIC methods is significantly more pronounced on the SIQAD dataset than on the SCID dataset, suggesting that the learning-based framework more effectively captures the features of web and office content, which dominates SIQAD dataset.
\begin{table}
  \begin{center}
	\caption{Ablation study.}
	\label{tab:ablation}
	\resizebox{0.8\linewidth}{!}{
	\begin{tabular}{cccc|c}
	\toprule
	\multicolumn{4}{c|}{ {Proposed modules} } & \multirow{2}{*}{BD-rate(\%)} \\
	\cmidrule{1-4}
	{MToRB} & {CTSFRB} & {MFCIM} & {AQ} & \\
	\midrule
	 {$\checkmark$} &  &  &  &   {0} \\ 
{$\checkmark$} & {$\checkmark$} &  & &  {-1.2} \\ 
{$\checkmark$} & {$\checkmark$} & {$\checkmark$} & &    {-4.3}\\ 
{$\checkmark$} & {$\checkmark$} & {$\checkmark$} & {$\checkmark$} &  {-5.9} \\
	\bottomrule
	\end{tabular}
	}
  \end{center}
\end{table}
\begin{figure}[htbp]
	\centering
	\begin{minipage}{0.45\textwidth}
	\centering
	\includegraphics[width=0.9\textwidth]{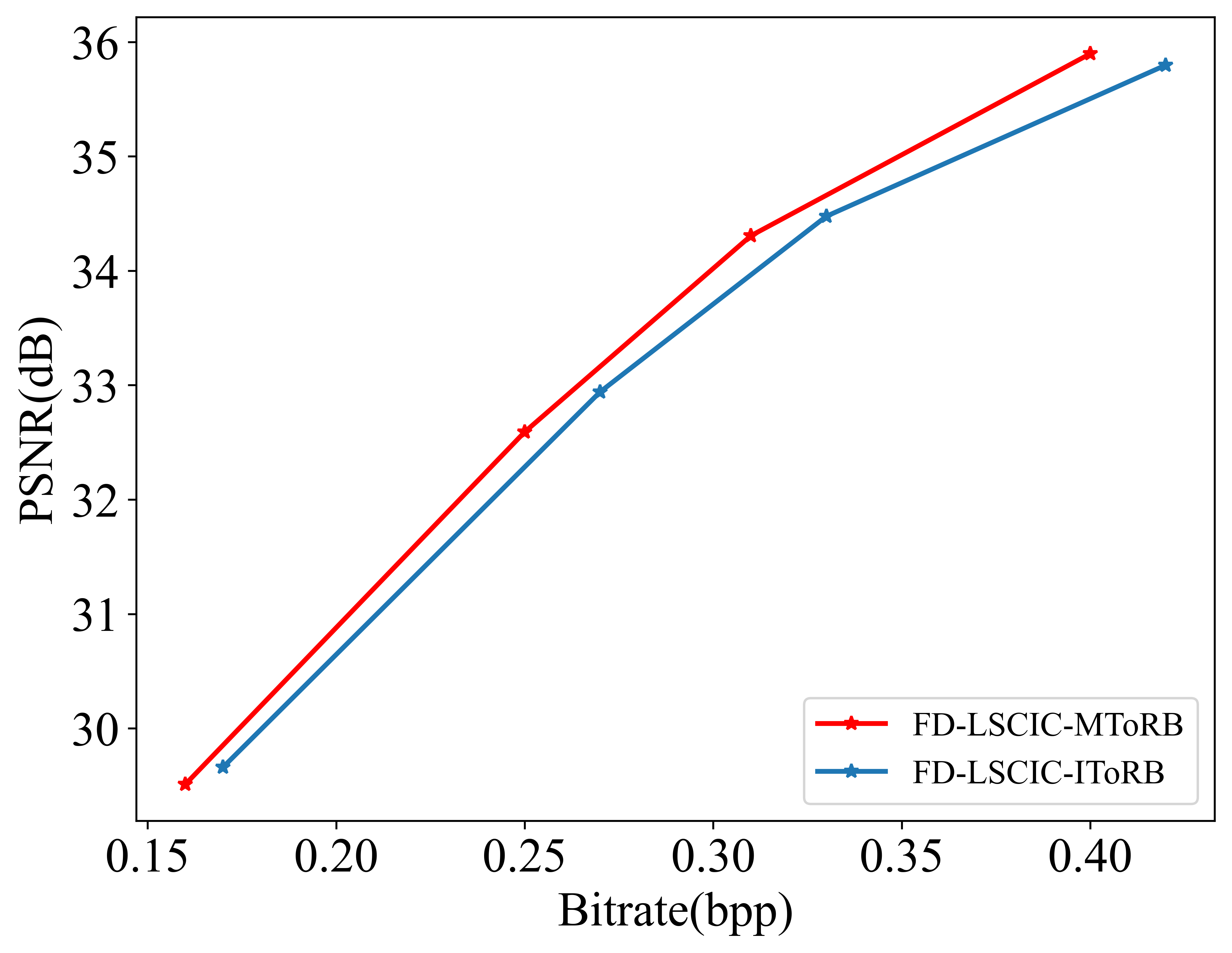}
	\caption{Ablation study of MToRB.}
	\label{fig:ablaiton-MToRB}
	\end{minipage}
\end{figure}
\vspace{-8pt}
\subsection{Complexity}
Table \ref{tab:bd-rate} also shows the time and computational complexity of different methods. Compared to H.266/VVC-SCC, the proposed method significantly reduces encoding time. Additionally, in contrast to methods \cite{cheng2020learned,jiang2024OMR,minnen2018joint} using serial context models, the proposed method employs parallel context models, resulting in noticeably faster encoding and decoding times.

\subsection{Qualitative results}
In Fig. \ref{fig:subject}, we illustrate the reconstruction quality of the codecs of SCI36 in the SCID dataset as an example. We can see that the proposed method outperforms BPG, H.266/VVC, H.266/VVC-SCC and OMR-net. Compared to other methods, our method achieves better reconstruction quality in terms of both PSNR and MS-SSIM.
\begin{figure}[t]
	\centering
	\begin{minipage}{0.45\textwidth}
	\centering
	\includegraphics[width=0.9\textwidth]{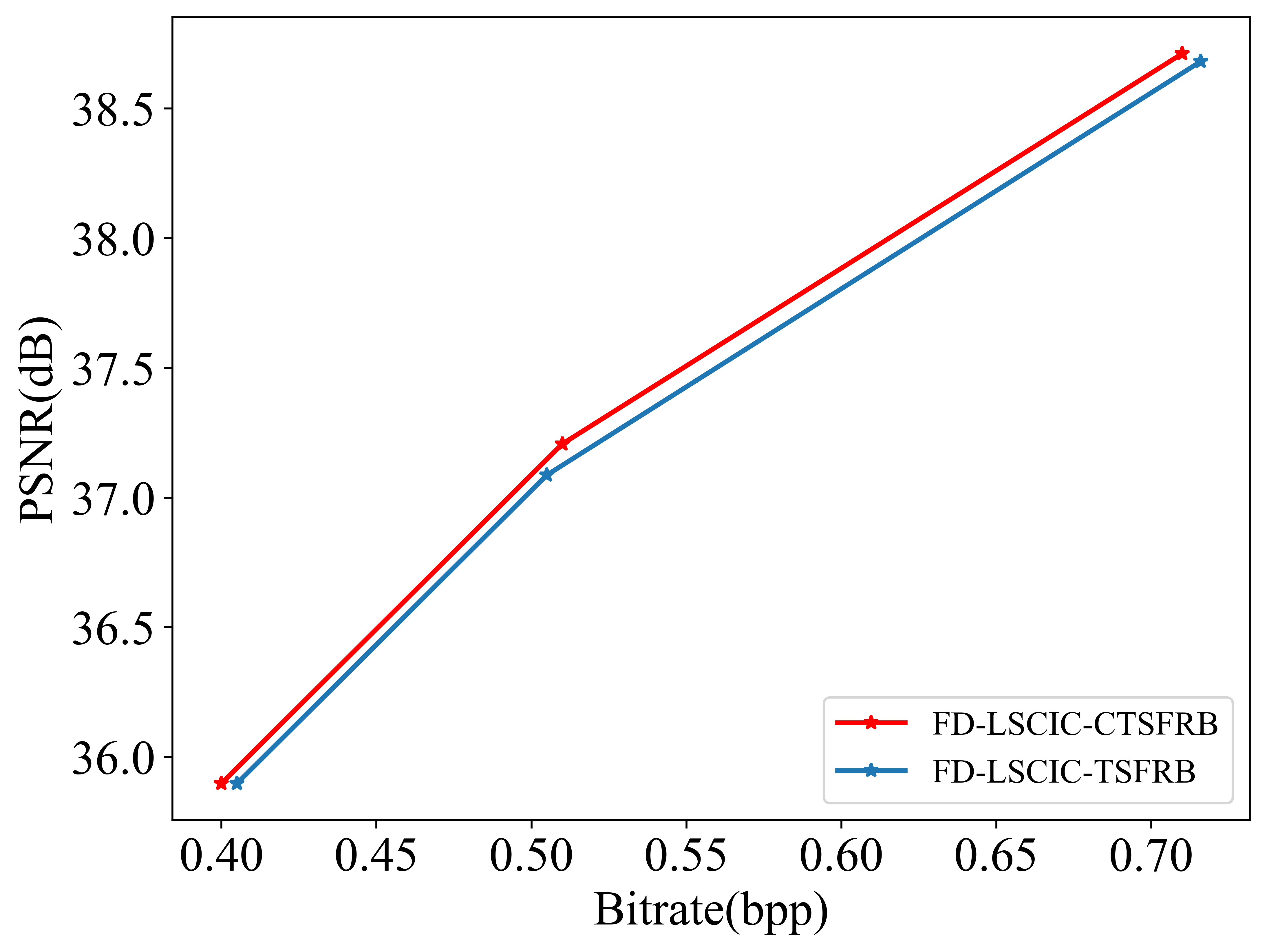}
	\caption{Ablation study of CTSFRB.}
	\label{fig:ablaiton-CTSFRB}
	\end{minipage}
\end{figure}
\subsection{Ablation Study}
\textbf{Performance of different components.} To validate the efficiency of the proposed components in the proposed method, we conducted ablation experiments on the SCID dataset, as shown in Table \ref{tab:ablation}. To calculate the BD-rate, we used four $\lambda$s (0.0035, 0.0067, 0.013, and 0.0483). 
Using the method with only MToRB as the anchor, adding CTSFRB reduced the BD-rate by 1\%, adding MFCIM further reduced it by 2\%, and incorporating AQ achieved an additional 2\% reduction. The results demonstrate that the proposed components can effectively enhance the RD performance.

\textbf{Performance of MToRB.} To validate the effectiveness of multi-frequency decomposition for SC image coding, we replaced MToRB with IToRB \cite{jiang2024OMR}, an octave convolution residual block that decomposes features into high and low frequencies. Fig. \ref{fig:ablaiton-MToRB} shows the RD performance on the SCID dataset. We can see that multi-frequency decomposition is more effective for SC images, which aligns with the findings presented in Fig. \ref{fig:Frequency characteristics}.
\begin{figure}[t]
\begin{minipage}[t]{0.48\linewidth}
  \centering
  \includegraphics[width=1\linewidth]{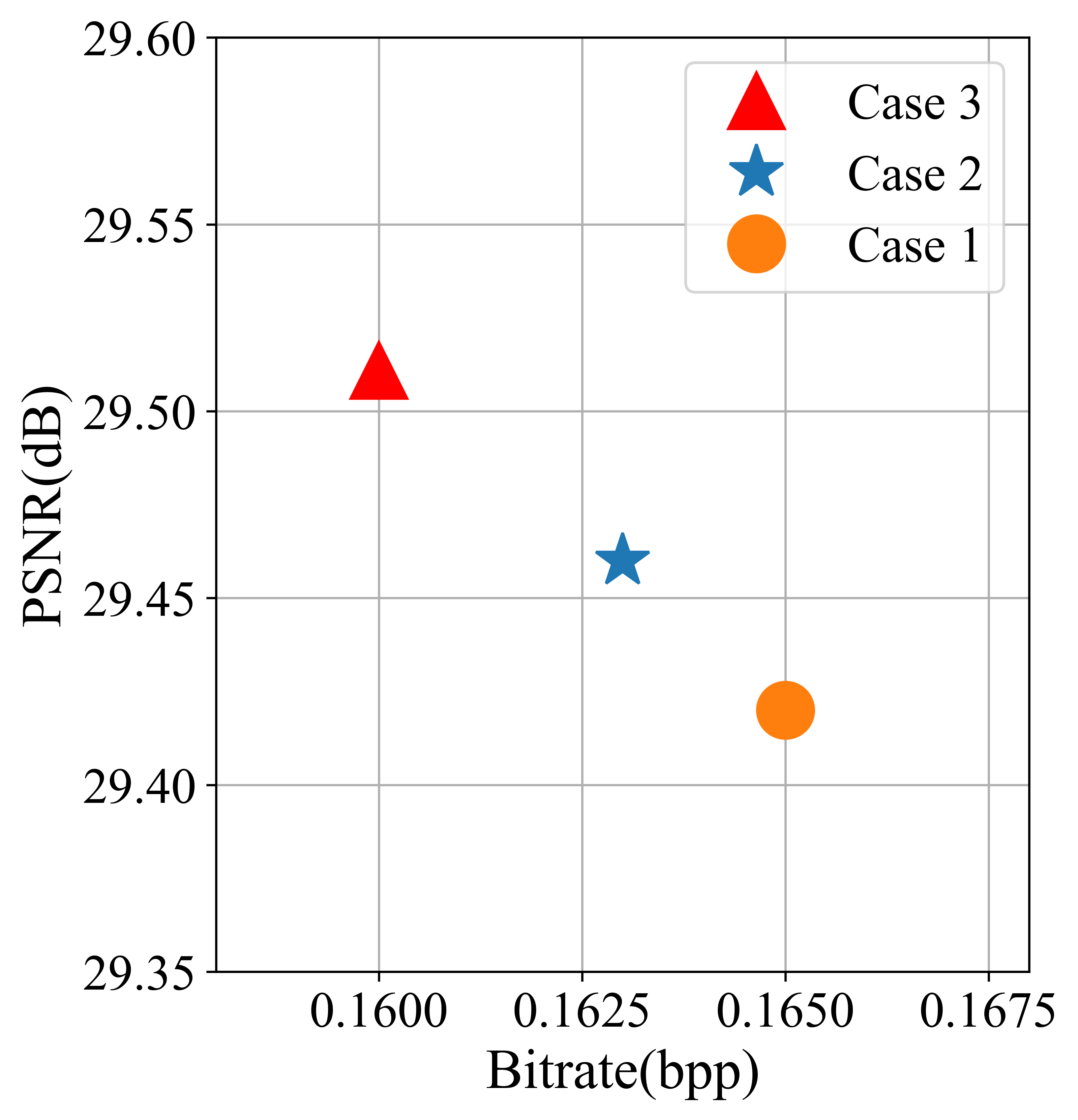}
  \centerline{(a) $\lambda$=0.0018}\medskip
\end{minipage}
\begin{minipage}[t]{0.48\linewidth}
  \centering
  \includegraphics[width=1\linewidth]{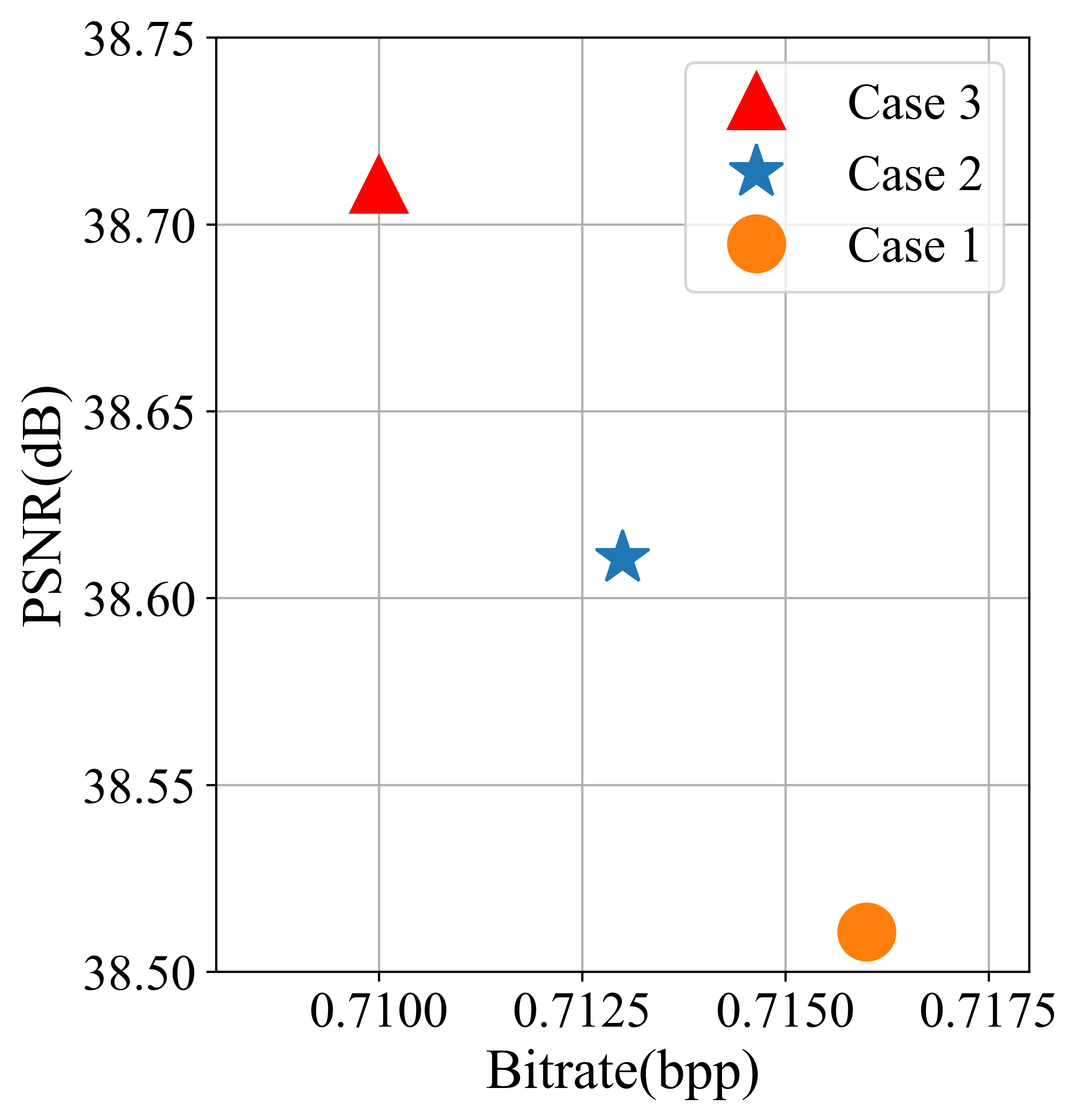}
  \centerline{(b) $\lambda$=0.0483}\medskip
\end{minipage}
\caption{Ablation study of MFCIM.}
\label{fig:abltion-MFCIM}
\end{figure}
\begin{figure}[t]
	\centering
	\begin{minipage}{0.49\textwidth}
	\centering
	\includegraphics[width=\textwidth]{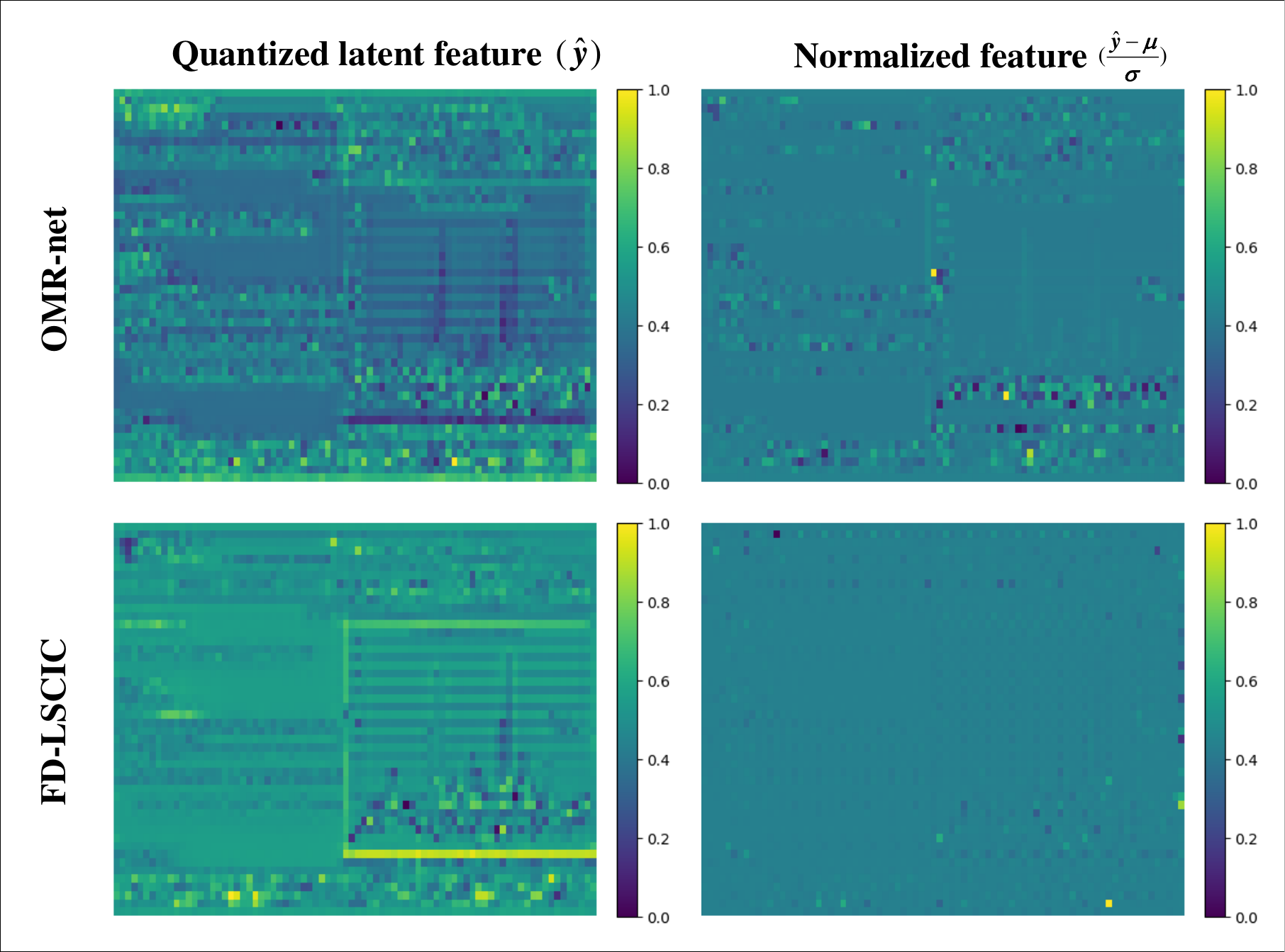}
	\caption{Visualization of latent feature map using SCI36 image from SCID dataset.}
	\label{fig:latent-visual}
	\end{minipage}
\end{figure}

\textbf{Performance of CTSFRB.} To verify the effectiveness of CTSFRB for SC image coding, we evaluated the RD performance using a single TSFRB and a CTSFRB consisting of two TSFRBs, as shown in Fig. \ref{fig:ablaiton-CTSFRB}. The results indicate that using CTSFRB further eliminates redundancy. Moreover, the improvement at higher bitrates is greater compared to lower bitrates.

\textbf{Performance of MFCIM.} Table \ref{tab:ablation} demonstrates the effectiveness of MFCIM for screen content image coding. To further explore the function of MFCIM, we designed three different context interaction cases. The first case (Case 1) utilizes only the high-to-low frequency context interaction network $C_{h-l}$. The second case (Case 2) adds the mid-to-low frequency context interaction module $C_{m-l}$ based on Case 1. The third case (Case 3) further includes $C_{h-m}$ based on Case 2. Fig. \ref{fig:abltion-MFCIM} illustrates the RD performance of these three cases. It can be seen that the approach incorporating $C_{h-l}$, $C_{m-l}$, and $C_{h-m}$ yields the best performance, indicating that context interaction across different frequency components effectively reduces feature correlation.

\textbf{Visualization of latent feature.} 
We compared the visualized feature map of the proposed method with OMR-Net \cite{jiang2024OMR}, as shown in Fig. \ref{fig:latent-visual} (left side). We can see that $\bm{\hat{y}}$ of the proposed FD-LSCIC is smoother than that of OMR-Net, indicating that FD-LSCIC can effectively address drastic frequency changes. From Fig. \ref{fig:latent-visual} (right side), FD-LSCIC achieves more uniform normalized latent representations, indicating that FD-LSCIC exhibits reduced residual redundancy compared to OMR-Net, thereby requiring fewer bits for encoding \cite{balle2018variational}.

\section{Conclusion}
We proposed a novel image compression method specifically designed for SC, addressing the unique characteristics and challenges associated with SC image coding. Our method introduces four modules to achieve multi-frequency feature extraction, fusion, context interaction, and adaptive quantization. In addition, we constructed currently the largest dataset for SC image compression, consisting of over 10,000 images from both PC and mobile platforms, covering typical screen content scenarios. Experimental results demonstrate that our approach outperforms traditional methods and existing LIC methods across different SC image datasets. In future, we aim to further explore the inter-frame correlation in SC to develop more efficient SC video compression schemes.

\bibliographystyle{IEEEtran}
\bibliography{ski}

\end{document}